\documentclass[final]{ustcstep}

\usepackage[dvips]{graphicx}
\usepackage[square]{natbib}

\def\gsim{\;\lower4pt\hbox{${\buildrel\displaystyle >\over\sim}$}\;}
\def\lsim{\;\lower4pt\hbox{${\buildrel\displaystyle <\over\sim}$}\;}
\def\grls{\;\lower4pt\hbox{${\buildrel\displaystyle >\over <}$}\;}

\newcommand\addr[2]{{\footnotesize \it $^{#1}$#2}\\}

\begin{document}

\title{Statistical Study of Coronal Mass Ejection Source Locations: Understanding CMEs Viewed in Coronagraphs}

\author{Yuming Wang, Caixia Chen, Bin Gui, Chenglong Shen, Pinzhong Ye, and S.
Wang\\[1pt]
\addr{}{KLBPP, School of Earth \& Space Sciences, University of
Science \& Technology of China, Hefei, Anhui 230026, China;}
\addr{}{Contact: ymwang@ustc.edu.cn}}

\maketitle
\tableofcontents

\begin{abstract}
How to properly understand coronal mass ejections (CMEs) viewed in white-light coronagraphs is
crucial to many relative researches in solar and space physics. The issue
is now particularly addressed in this paper through studying the source locations of all the
1078 LASCO CMEs listed in CDAW CME catalog during 1997 -- 1998 and their correlation with
CMEs' apparent parameters. By manually checking LASCO and EIT movies of these
CMEs, we find that, except 231 CMEs whose source locations can not be identified due to
poor data, there are 288 CMEs with location identified on the front-side solar disk,
234 CMEs appearing above solar limb, and 325 CMEs without evident eruptive signatures in the
field of view of EIT. Based on the statistical results of CMEs' source locations, four physical
issues, including (1) the missing rate of CMEs by SOHO LASCO and EIT, (2) the mass of CMEs, (3)
the causes of halo CMEs and (4) the deflections of CMEs in the corona, are exhaustively analyzed.
It is found that (1) about 32\% of front-side CMEs can not be recognized by SOHO, (2) the brightness
of a CME at any heliocentric distance is roughly positively correlated with its speed, and the CME
mass derived from the brightness is probably overestimated, (3) both projection effect and violent
eruption are the major causes of halo CMEs, and especially for limb halo CMEs, the latter is the
primary one, (4) most CMEs deflected towards equator near the solar minimum, and these deflections
can be classified into three types, the asymmetrical expansion, non-radial ejection, and the deflected
propagation.
\end{abstract}


\section{Introduction}

Coronal mass ejections (CMEs) are recognized as transient bright features in the field of view
(FOV) of white-light coronagraphs. However, their apparent properties/behaviors
manifested in coronagraphs may not reflect what the CMEs actually should be,
as observations of coronagraphs have at least three intrinsic limitations.
The first one comes from the projection effect. All the three-dimensional
information is embedded in two-dimensional images. Thus the position or
speed of a CME measured in coronagraphs is only the projection of real
position or speed on the plane of the sky, the shape of a CME
depends on the angle of view, and the brightness recorded
is an integral of the photons scattered by free electrons along the
line-of-sight. The second one, we called occulting effect, is due to the
occulting disk, which is used by coronagraphs to block the photons directly emitted from
the photosphere. It was clearly pointed out by \citet{Howard_etal_1982}
that two identical CMEs originating from the solar limb and
disk-center, respectively, will look much different. The time and
heliocentric distance of the disk-center CME entering the FOV of a coronagraph
will be later and farther than those of the limb CME. It will further cause the disk-center CME
fainter and diffuser than the limb CME.
The third one is because of the
Thomson scattering effect \citep[e.g.,][]{Hundhausen_1993, Andrews_2002,
Vourlidas_Howard_2006}. This effect results in a so-called Thomson sphere,
on which the plasma material is the most visible.

Moreover, in most popular coronagraph images, the inner corona is hidden
behind the occulting disk. For example, the occulting disk size of the coronagraph
LASCO/C2 onboard the SOHO spacecraft is 2 $R_S$, and it is 1.4 $R_S$ for the
coronagraph COR1 onboard the STEREO twin spacecraft. Thus, we are blind to the
CME behavior in the region covered by the occulting disk,
where the CME propagation trajectory may change significantly.
Here, we use the term `deflection' for the behavior of CME's non-radial
ejection and/or propagation. It is an important factor for space weather.
As early as \citeyear{MacQueen_etal_1986},
\citeauthor{MacQueen_etal_1986} had found the CME deflections
in latitudinal direction by measuring 29 CMEs observed by
the Skylab. \citet{Gopalswamy_etal_2000a} discussed the non-radial
propagation of the 1997 December 14 CME and pointed out that such a
phenomenon clearly implied the constraint of the complex multi-polar
structures surrounding the CME \citep{Webb_etal_1997,
Gopalswamy_etal_2004}. With more CME events detected by LASCO during
1996 to 2002, {\it Cremades} and coworkers carried out a statistical
study on their defined `structured' CMEs. They found that many CMEs do not propagate radially with
respect to their source locations, and the neighboring and/or polar
coronal holes played a major role in causing the deflections of CMEs
\citep{Cremades_Bothmer_2004, Cremades_etal_2006}.

The presence of these effects requires us to be very careful when we
interpret the observed bright features in coronagraphs. Only white-light
images from coronagraphs are not enough. The information of the solar
source locations of all CMEs is necessary. There have been
some efforts except for the previously mentioned work about CME deflections.
\citet{Yashiro_etal_2005} investigated 1301 X-ray flares with
intensity larger than C3 and their associations with CMEs, and found that
about 14\% of white-light CMEs were missed by LASCO. The statistically study
of 9224 LASCO CMEs from 1996 to 2004 by \citet{Lara_etal_2006} suggested
that halo CMEs are different from normal CMEs, which can not be merely
explained by projection effect, and the brightness of halo CMEs probably
includes their driven (shock) waves.

We acknowledge that these previous studies have advanced our understanding
of the white-light CMEs observed by coronagraphs, but it is not comprehensive. We also
realize that there are few works identifying the source locations of all
CMEs no matter whether the CME is halo or narrow, strong or faint.
Most studies involving the information of source
locations considered halo CMEs only \citep[e.g.,][]{Wang_etal_2002a,
Zhou_etal_2003, Zhao_Webb_2003}. Some others set certain criteria in
the selection of CMEs. For example, the study by
\citet{Subramanian_Dere_2001} only included the 32 CMEs with very
clear EUV signatures on the solar surface.
\citet{Cremades_Bothmer_2004} selected so called `structured' CMEs,
in which halo, narrow or faint CMEs are all
excluded. \citet{Yashiro_etal_2005} work involved the CMEs
associated with flares above C3 level. To our knowledge, the study by
\citet{Plunkett_etal_2001}, might be the only statistical work, in
which all the CMEs during the period of interest, which is from 1997
April to December, were identified for their source locations.

An incomplete or biased sample may lead to unreliable or one-sided results,
particularly, based on observations of coronagraphs, which have some
intrinsic limitations. In this paper, we will identify the source locations
of all the 1078 CMEs from 1997 to 1998 listed in the CDAW CME
catalog\footnote{A widely-used manually-compiled catalog, refer to
\url{http://cdaw.gsfc.nasa.gov/CME_list/}} \citep{Yashiro_etal_2004}, and try to better understand
CMEs viewed in white-light coronagraphs. Except for the statistical results
of CMEs' source locations, our investigation will address the
following four issues.
\begin{enumerate}
\item {\it Missing rate of CMEs.} How many CMEs were missed by LASCO and EIT, and how
many front-side CMEs were unnoticed by SOHO?

\item {\it Mass of CMEs.} Whether or not can the enhanced brightness in
coronagraphs reflect the CME mass?

\item {\it Causes of halo CMEs.} Why do some CMEs manifest a halo appearance?

\item {\it Deflections of CMEs.} How often and significant are CMEs deflected in the corona and why?
\end{enumerate}

The period of 1997 -- 1998 is the beginning of the ascending phase of
solar cycle 23, during which the solar condition is relatively simple
and the solar activity level is low. Thus the source locations of CMEs are
relatively easy to be identified with small ambiguity. This paper is
organized as follows. In the next section, we present our data source, and
particularly focus on the identification and classification of the CME source locations.
The statistical results of the CME source locations are shown in Sec.\ref{sec_source}.
In Sec.\ref{sec_implications}, the four issues mention above are extensively
discussed. A summary and conclusions are given in Sec.\ref{sec_conclusions}.

\section{Data Preparation}\label{sec_data}

The CDAW CME catalog provides so far the most reliable list of CMEs
recorded by SOHO/LASCO, in which some CME apparent parameters, such as
angular width, position angle, linear speed, etc, are included. Since
only the LASCO data are used by the catalog, there is no information of
CMEs' source locations. To identify the source locations of CMEs, the
SOHO/EIT 195 \AA\ images are used. The identification method is similar
to that employed by \citet{Wang_etal_2002a}, in which the time and propagation
direction of a CME obtained from the LASCO movie is used to roughly locate
the time and region of the CME in EIT 195 \AA\ images and then this region
is carefully checked if there is any EUV eruptive activity associated with
the CME.

Lots of observations have suggested that various
eruptive activities appearing at various wavelengths on the solar surface
probably indicate the launch
of a CME. These signatures could be flares in multiple wavelengths,
dimmings and waves in EUV passbands, post-eruptive loops/arcades in
X-ray and EUV images, etc. However, a CME process
may not be companied with all of these phenomena. Flares are
thought to be tightly related with CMEs \citep[e.g.,][]{Harrison_1995,
Harrison_2003, Zhang_etal_2001a}, but it has been statistically
suggested that flares are not one-to-one associated with CMEs,
vice versa \citep[e.g.,][]{StCyr_Webb_1991, Wang_etal_2002a,
Zhou_etal_2003, Andrews_2003, Yashiro_etal_2005, Yermolaev_Yermolaev_2006}.
A flare even stronger than X class could be associated without a CME
\citep{Green_etal_2002, Yashiro_etal_2005, Wang_Zhang_2007}. A more
confident solar surface signature of a CME is the combination of a
flare and EUV dimming and/or waves. Thus, in our identification procedure,
we assume that such a combined signature in EIT 195 \AA\ images indicates
a CME originating from visible solar disk.

Meanwhile, we realize that there is no conclusion that a front-side CME
must be accompanied with some visible EUV signature on the solar surface,
which was emphasized by \citet{Yermolaev_Yermolaev_2006} and \citet{Yermolaev_2008}.
It means that a CME without any eruptive signature in EIT 195 \AA\ images
might come from front-side solar disk. As will be discussed in
Sec.\ref{sec_invisible}, such CMEs do exists. This has also been noted
in the following classifications.

\begin{figure*}[tbh]
  \centering
  \includegraphics[width=\hsize]{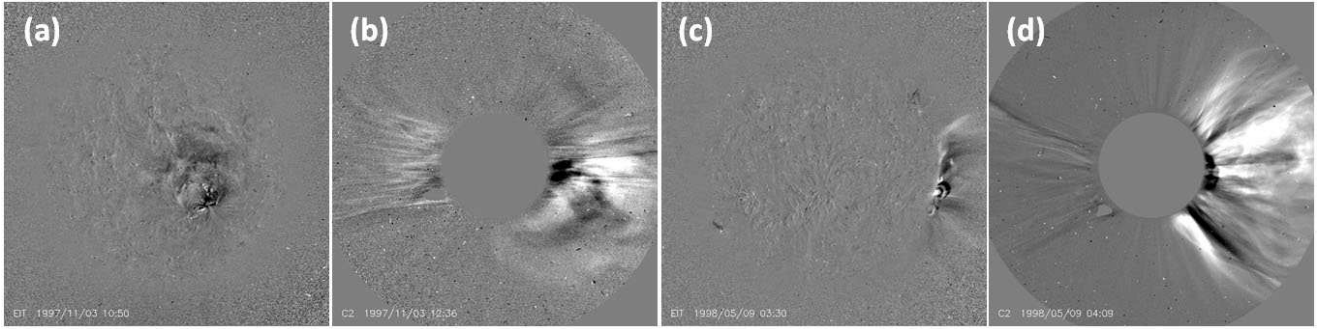}
  \caption{Panel (a) and (b) show the EIT and LASCO images of a LI CME on 1997 November 3.
The CME's source location can be identified on the visible solar disk. Panel (c) and (d)
show the same images of an AL CME on 1998 May 9, whose eruptive signature can only be seen
above the west limb.}\label{fg_classification}
\end{figure*}

\begin{figure*}[tbh]
  \centering
  \includegraphics[width=\hsize]{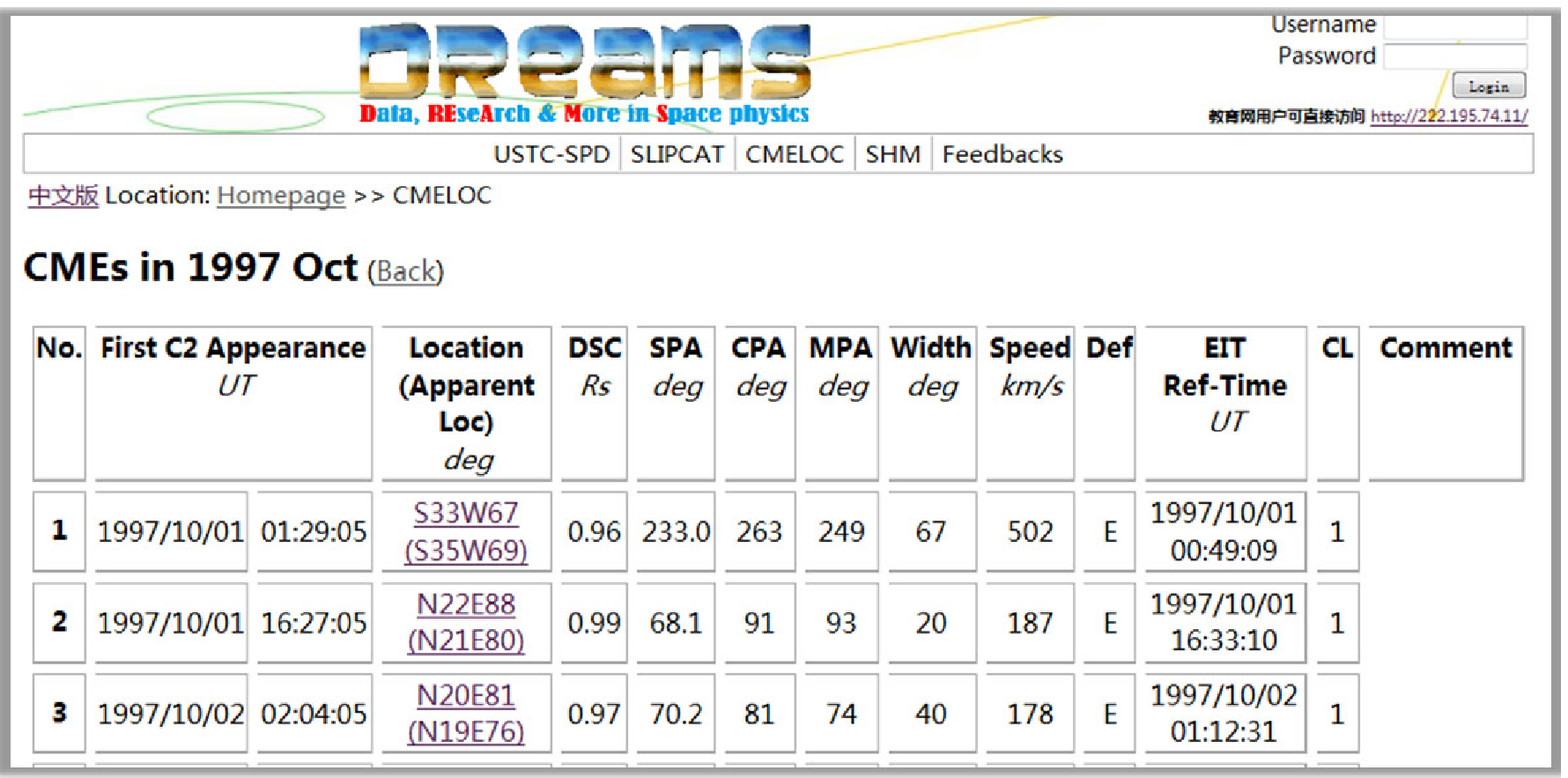}
  \caption{A snapshot of the web-based on-line list of CMEs. Visit {http://space.ustc.edu.cn/dreams/cme\_sources/}
for details.}\label{fg_web}
\end{figure*}

After a manually check of the EIT observations, the source locations of
all the 1078 CMEs during 1997 -- 1998 are identified. It is found that
these CME source locations can be classified as the following four subsets.

\emph{1. Location Identified (LI):} For a CME in this subset,
evident eruption features, such as brightening and/or dimming, on
the solar disk seen in EIT 195 \AA\ can be related to it. Such a
CME definitely originated from the front-side of the Sun.
Figure~\ref{fg_classification}(a) and \ref{fg_classification}(b)
show an example CME viewed in EIT and LASCO. From the EIT 195 \AA\
image, we could assign a location (given in latitude and longitude)
for the CME, which is usually the center of the eruption feature and
the error is about 5 -- 10 degrees. The measured location coordinates
directly from EIT images are the apparent coordinates, but not in the
heliographic coordinates. The heliographic coordinates can be calculated
by applying the correction of the angle between the solar equatorial
plane and ecliptic plane. According to each pair of the apparent
coordinates, we can further derive the following two parameters.
One is the projected distance of the source location from disk
center ($DSC$) in the plane of sky. The other is the position
angle of the source location ($SPA$). These two parameters are
useful in the analysis of the projection effect, visibility and deflections
of CMEs.

\emph{2. Above Limb (AL):} In this subset, we can only find eruption
features associated with CMEs mainly above the solar limb as illustrated by
the example shown in Figure~\ref{fg_classification}(c) and
\ref{fg_classification}(d). We could expect that these CMEs
probably originated from the backside and were close to the limb of the Sun. For
such a CME, the parameter $DSC$ can not be obtained, but $SPA$ could
still be roughly estimated from the EIT images.

\emph{3. No Signature (NS):} No any eruption features were seen in
the FOV of EIT for this subset of CMEs. Such CMEs probably
originated from the backside of the Sun. Also it is possible that
some of them launched from the front-side solar disk but had very
weak signatures or originated at an altitude not corresponding to
the EUV 195 \AA\ passband.

\emph{4. Poor Data (PD):} The source locations of these CMEs can
not be identified, because of low cadence,
unqualified images and/or data gaps in EIT 195 \AA\ data.

Table~\ref{tb_classification} lists the numbers of CMEs for the first
three subsets. Except 231 PD CMEs that we have no sufficient data to
identify their source regions, a total of 847 CMEs have been checked
carefully, and it is found that there are 288 (occupying about 34\%)
CMEs with the source location at front-side solar disk, 234 ($\sim28\%$) CMEs
having been found above limb, and 325 ($\sim38\%$) CMEs without any eruptive signatures in
EUV 195 \AA\ passband.
Meanwhile, we give the confidence level ($CL$) of identification.
Three levels are given. Level 1 means that the identification is
confident; level 3 means ambiguous; and level 2 is between.

\begin{table}[tbh]
\begin{center}
\caption{CME Numbers in the Different Subsets} \label{tb_classification}
\begin{tabular}{cccccc}
\hline
CL & 1 &2 &3 & Subtotal & Percentage \\
\hline
LI   & 189 & 60 & 39 & 288 & 34\%\\
AL   & 160 & 60 & 14 & 234 & 28\%\\
NS   & 214 & 92 & 19 & 325 & 38\%\\
\hline
Total   & 563 & 212 & 72 & 847 &100\%\\
\hline
\end{tabular}
\end{center}
\end{table}

A list containing the information of the CMEs' source regions has
been compiled at the website
\url{http://space.ustc.edu.cn/dreams/cme_sources/}.
Figure~\ref{fg_web} is a glance of the list. For
each CME, the list integrates the parameters from CDAW CME catalog
($CPA$ and $MPA$, angular width, linear speed, etc.) and our
own parameters (source location, $DSC$, $SPA$, $CL$, etc.). One can visit the website for more
details. If not otherwise specified, in the following analysis,
we only include the LI CMEs with $CL$ equal to 1 and 2, which count
the number of 249.

\section{Statistical Results of Source Locations}\label{sec_source}
\subsection{Distribution of CME Source Locations}\label{sec_dis}
\begin{figure}[tb]
  \centering
  \includegraphics[width=\hsize]{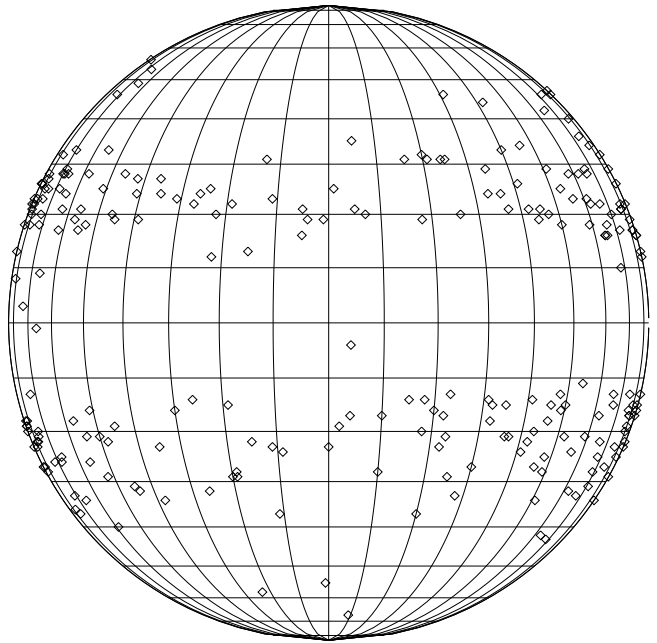}
  \caption{Distribution of the source locations of CMEs on a meshed solar disk.}\label{fg_solardisk}
\end{figure}

Figure~\ref{fg_solardisk} shows the distribution of the source
locations of the LI CMEs. Some quick results could be obtained immediately. About
52\%/48\% of CMEs launched from northern/southern hemisphere, and about
54\%/46\% of CMEs originated from western/eastern hemisphere. Further,
we consider a CME with $DSC$ equal or larger than 0.85 $R_S$ as a limb
event (otherwise an on-disk event) and a CME with angular width larger
than 100$^\circ$ as a halo event (otherwise a non-halo event). It is
found that about 56\% of CMEs come from solar limb
(comparing with 44\% of on-disk CMEs), and 18\% of CMEs are halo
(comparing with 82\% of non-halo CMEs). Table~\ref{tb_sources}
summarizes the numbers of the CMEs.

\begin{table*}[tbh]
\begin{center}
\caption{Numbers of Different Kinds of LI CMEs} \label{tb_sources}
\begin{tabular}{c|c|cc|cc|cc|cc}
\hline
&Total & Northern &Southern &Western &Eastern & On-disk &Limb &Halo &Non-halo  \\
\hline
Number & 249 &129 &120 &135 &114 &110 &139 &44 &205 \\
Percentage &100\% &52\% &48\% &54\% &46\% &44\% &56\% &18\% &82\% \\
\hline
\end{tabular}\\
{See the text for the definitions of the terms on-disk,
limb, halo and non-halo.}
\end{center}
\end{table*}

\begin{figure*}[tb]
  \centering
  \includegraphics[width=0.495\hsize]{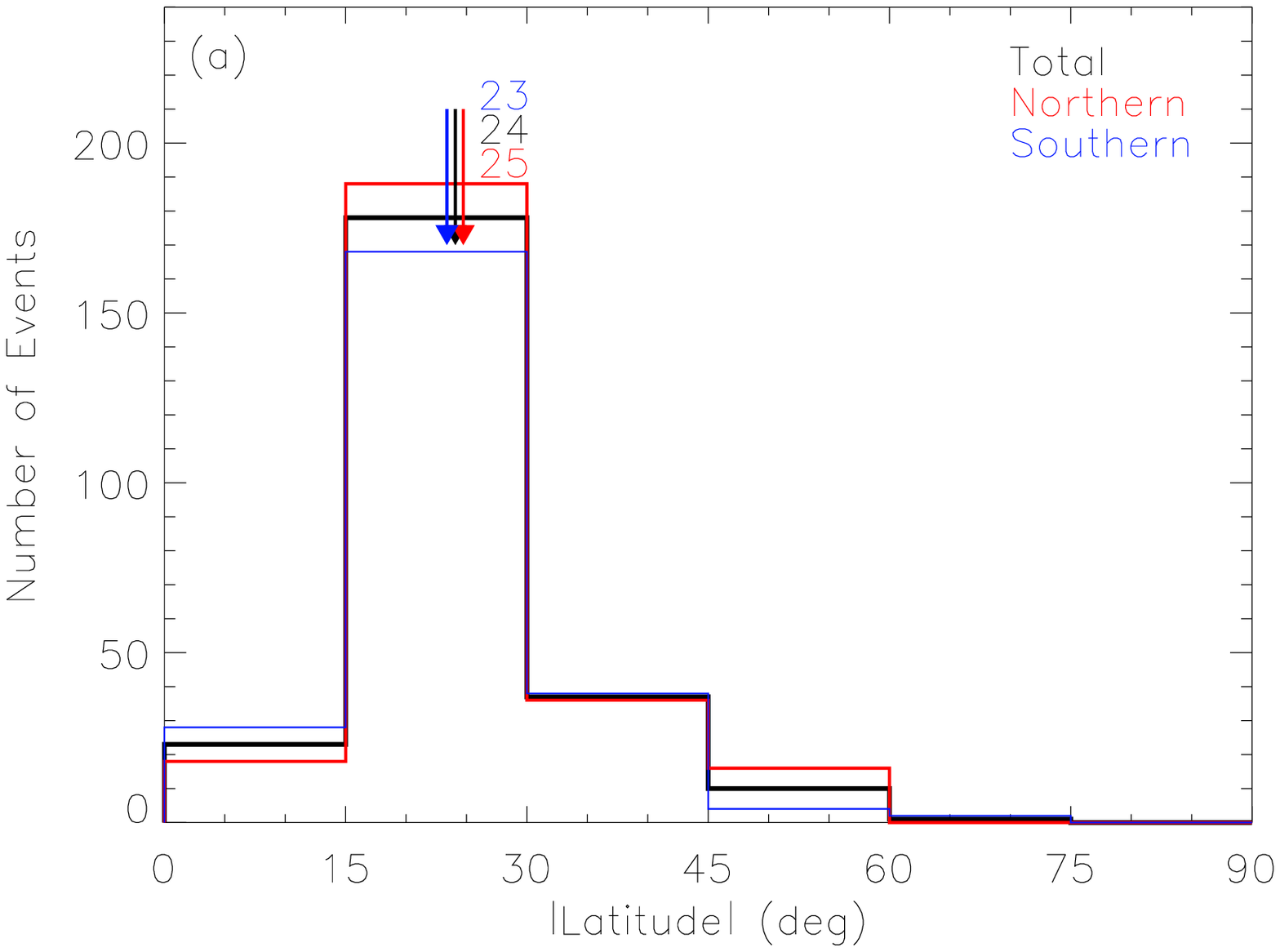}
  \includegraphics[width=0.495\hsize]{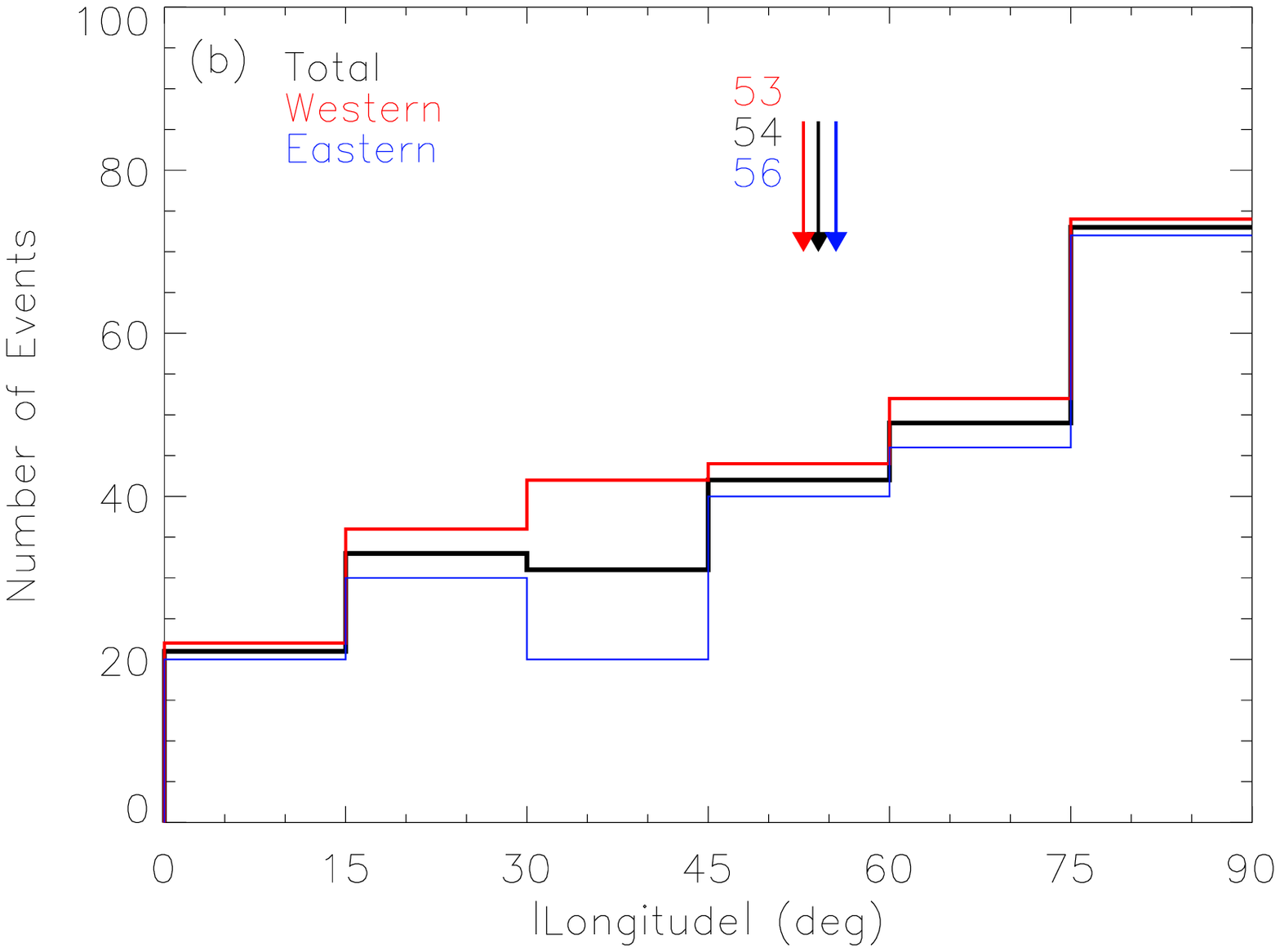}
  \caption{Histograms of the absolute values of the latitude and longitude for
all CMEs and the CMEs in other two subsets (northern/southern for latitude and
east/west for longitude). For clarity, the CME counts in the subsets are multiplied
by a factor of 2. The color-coded arrows and numbers indicate the average values.}\label{fg_dsl}
\end{figure*}

Figure~\ref{fg_dsl}(a) shows the distribution of the CMEs' source
locations in latitude. The black line presents all the 249 CMEs, the
red line is for the northern CMEs, and the blue line for southern
CMEs. Note the numbers of the northern and southern CMEs are multiplied
by a factor of 2 for clarity. Obviously, the distribution is south-north symmetrical
and has a clear bimodal appearance with two
outstanding peaks locating in $\pm(15^\circ-30^\circ)$, respectively. The average
latitude is $\sim\pm24^\circ$, and it can be estimated that $\sim71\%$ of CMEs
originated from $\pm(15^\circ-30^\circ)$.
Moreover, there is no CME originating beyond $\pm75^\circ$. The bimodal distribution
is different from the distribution of CMEs' apparent latitudes measured in LASCO
images, which is a distribution with only one peak near the solar equator \citep[e.g.,][]{StCyr_etal_2000,
Yashiro_etal_2004}. Such a difference was pointed out by \citet{Plunkett_etal_2001}.
The reason why the distribution of the latitudes of source locations differs from
that of the apparent latitudes could be (1) projection effect \citep{Hundhausen_1993},
and (2) that most CMEs may not eject/propagate radially, but undergo an
equator-ward deflection (refer to Sec.\ref{sec_deflection}). The first
reason can be seen from the work by \citet{Burkepile_etal_2004}, who studied
the 111 limb CMEs observed by SMM and found a similar bimodal distribution
of the CMEs' apparent latitudes with peaks at about $\pm15^\circ$.

The longitude distribution of the CMEs' source locations is
presented in Figure~\ref{fg_dsl}(b). Similarly, there is no
east-west asymmetry.
The CME count is not uniformly distributed along the
longitude, but increases with the increasing absolute longitude.
The average longitude is
$\sim\pm54^\circ$, and about 49\% of CMEs originated from the
regions outside $\pm60^\circ$. The non-uniform distribution of the
longitude suggests that the CMEs originating from solar limb could
be observed more easily than those near the disk center. The
visibility of CMEs has been studied before
\citep[e.g.,][]{Yashiro_etal_2005}. The three intrinsic limitations
of coronagraph observations mentioned in the section of Introduction
are responsible for such phenomenon. A more detailed discussion of
the CME visibility or missing rate of CMEs will be given in the next section.

\begin{figure}[tb]
  \centering
  \includegraphics[width=\hsize]{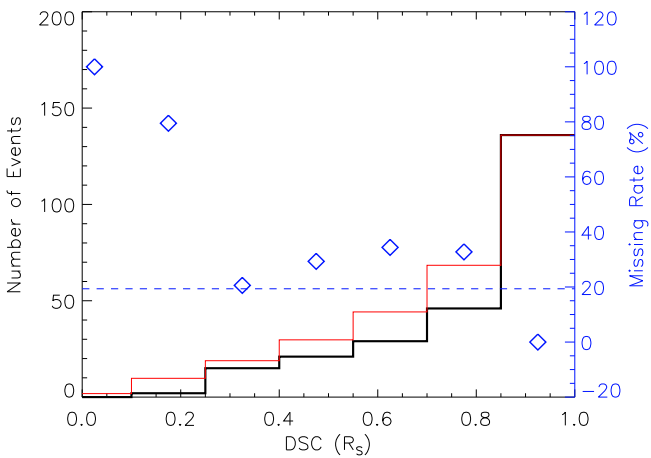}
  \caption{Histogram of $DSC$. The black one presents the observed CME
counts, and the red one the expected CME counts. The missing rate of CMEs
is given by the blue diamond symbols, which is measured by the vertical axis on the right. The dashed line marks the average missing rate. (See text for details)}\label{fg_dsc}
\end{figure}

Combine the information of both latitude and longitude, we can get the
distribution of CME counts with respect to the parameter $DSC$, as presented
by the black histogram in Figure~\ref{fg_dsc}. It is found that the CME
count increases dramatically as $DSC$ increases.
About 56\% of CMEs took place outside of $DSC=0.85R_S$, namely limb CMEs.

\subsection{Correlations between CME Source Locations and Apparent Properties}\label{sec_apparent}
\begin{figure*}[tb]
  \centering
  \includegraphics[width=0.495\hsize]{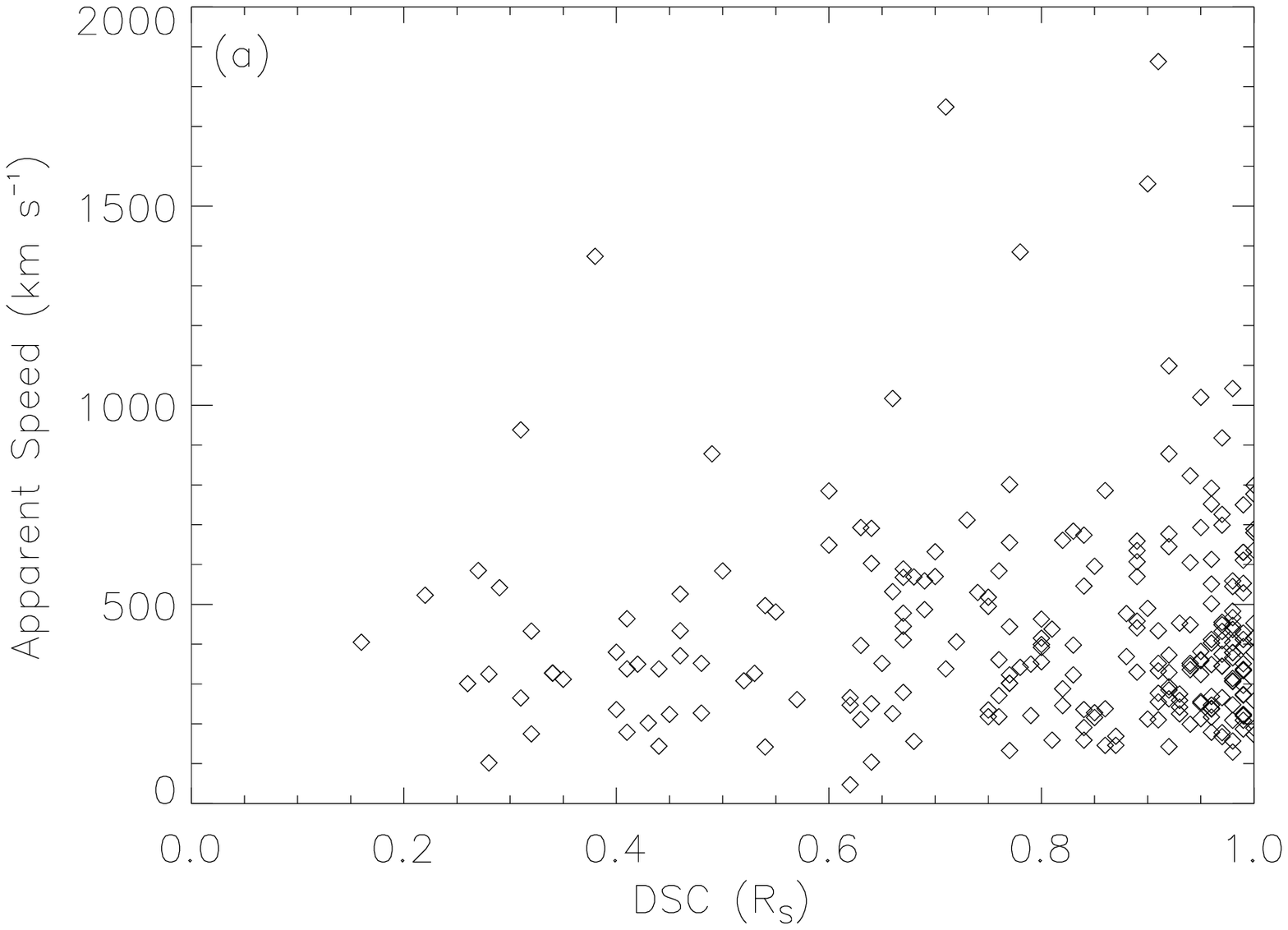}
  \includegraphics[width=0.495\hsize]{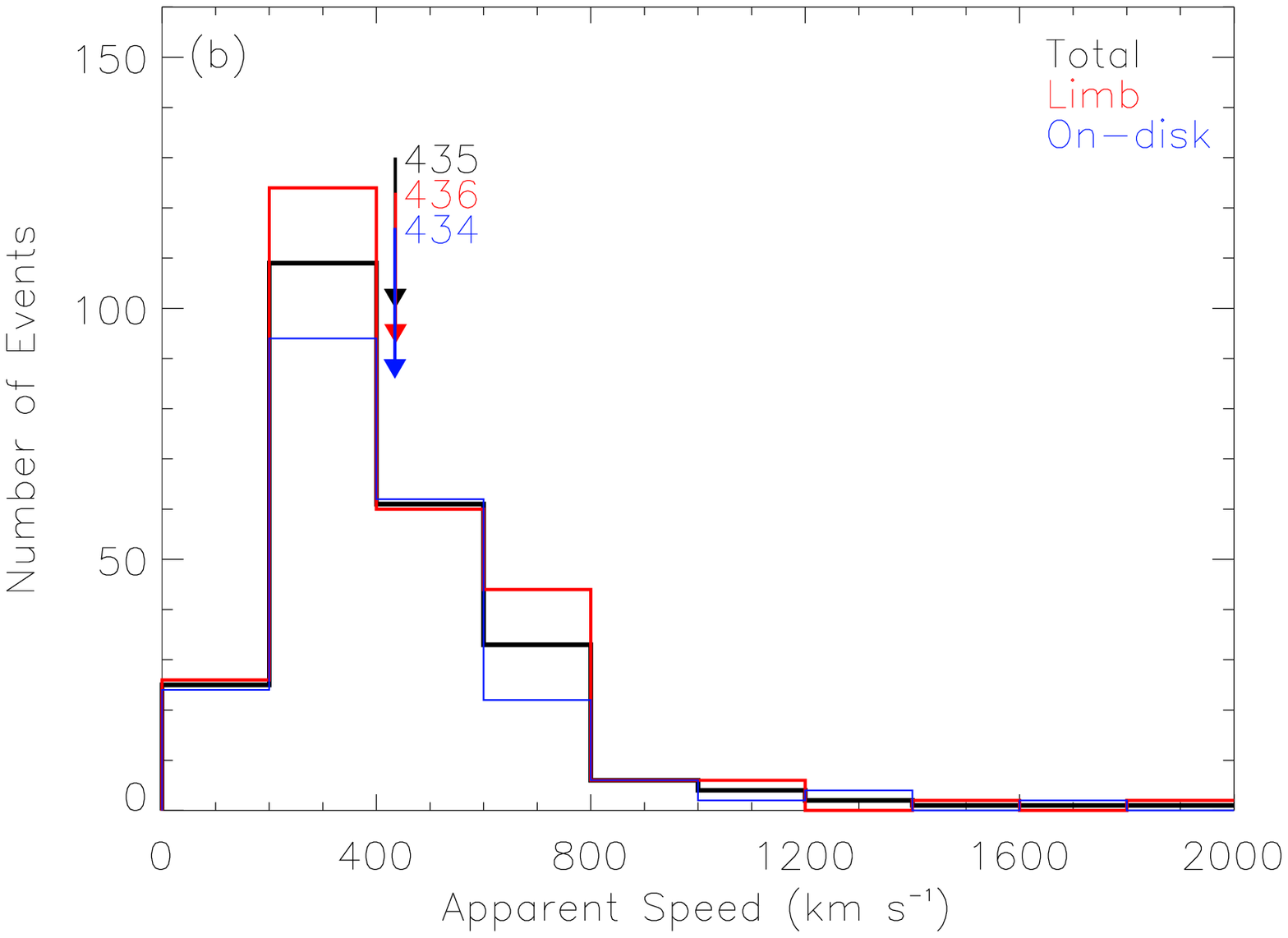}
  \includegraphics[width=0.495\hsize]{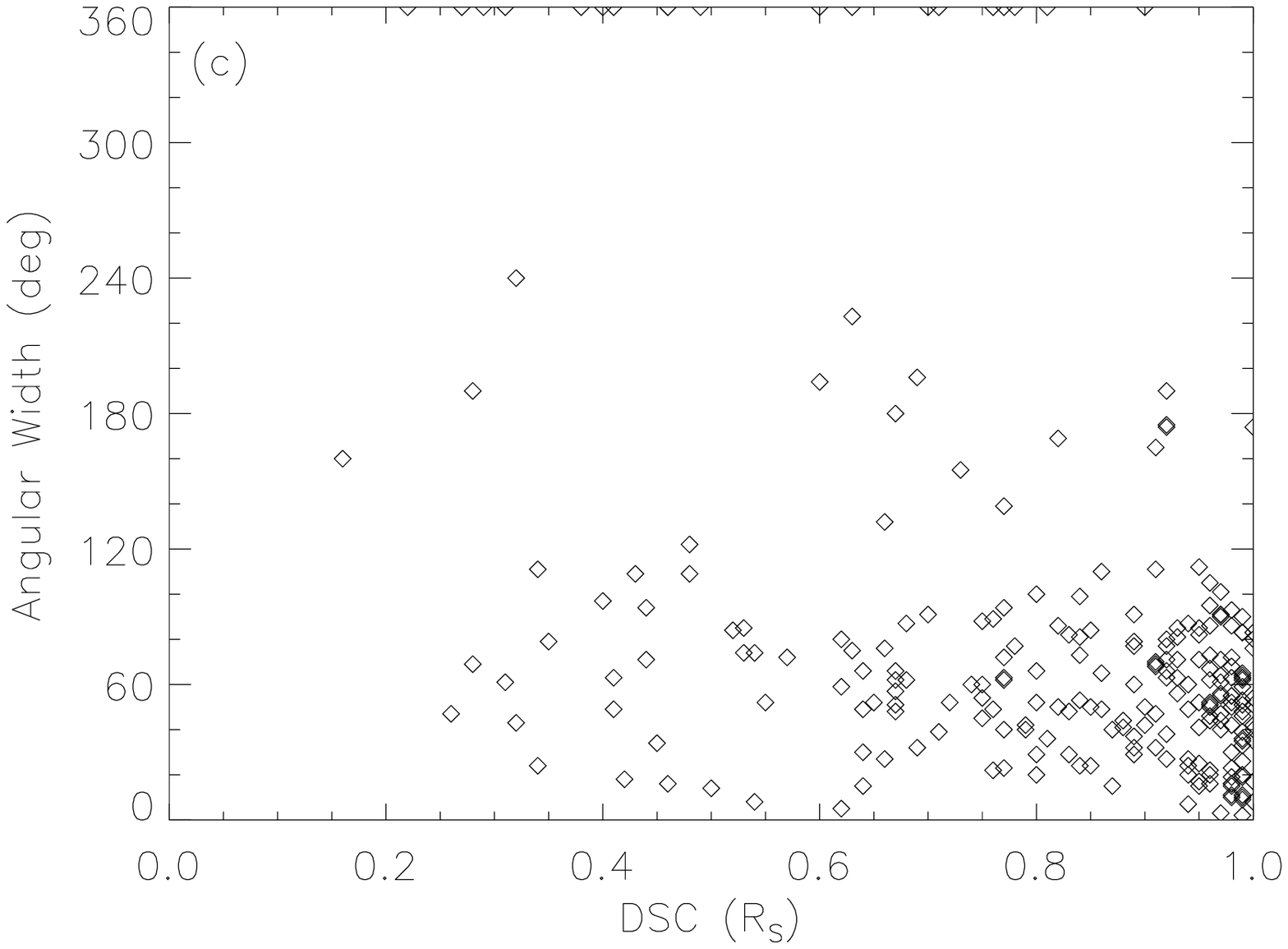}
  \includegraphics[width=0.495\hsize]{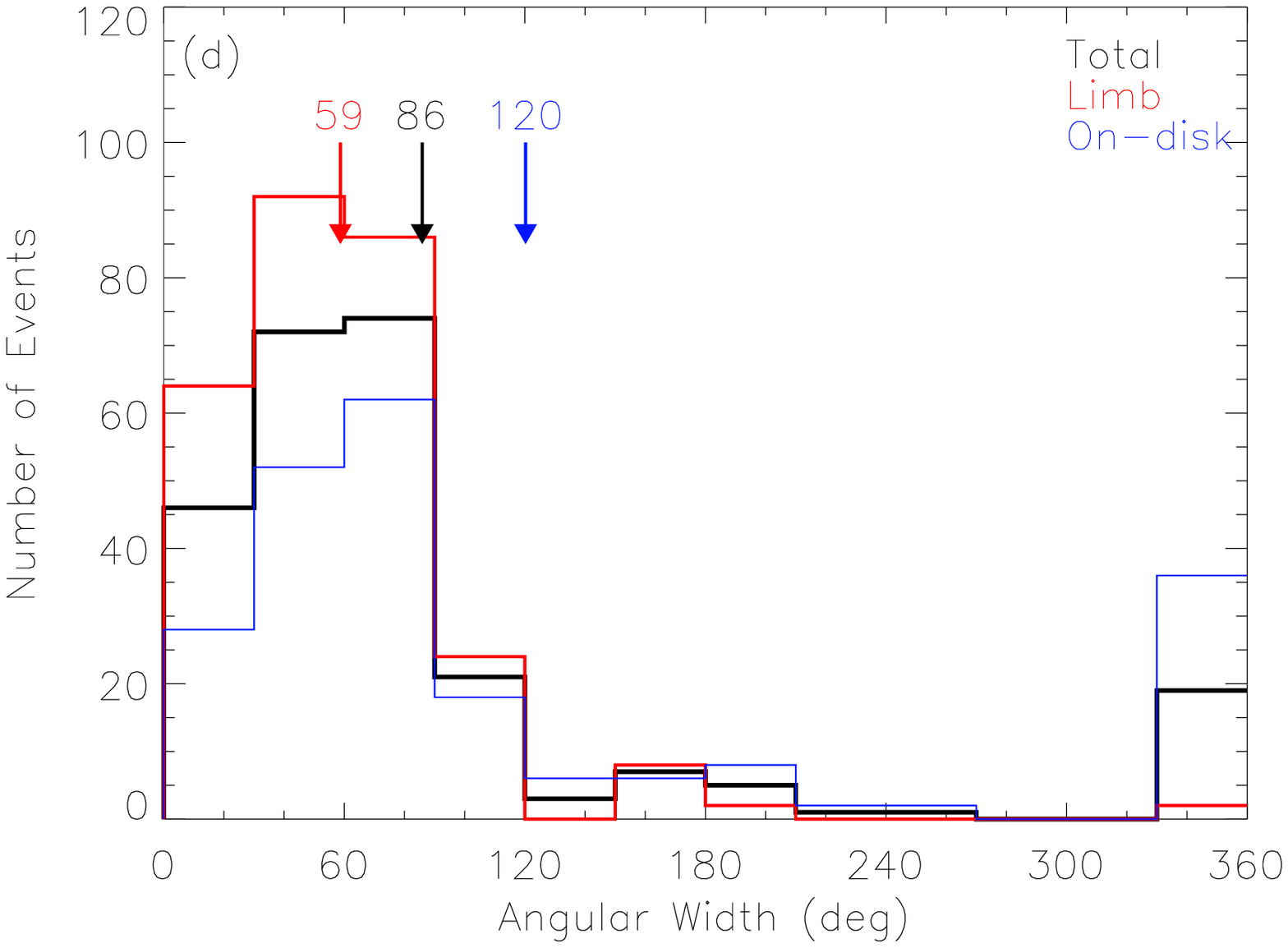}
  \caption{Panel (a) and (c) show the apparent speed and angular width, respectively, of CMEs
as a function of $DSC$. Panel (b) and (d) present the histograms of the two parameters with the pattern
as same as that in Figure \ref{fg_dsl}.}\label{fg_speed}
\end{figure*}

Figure~\ref{fg_speed}(a) and \ref{fg_speed}(b) display the
scattering plot of the CME apparent speeds versus source locations, and the
histograms of the speeds for limb, on-disk and all CMEs,
respectively. The both plots do not show any evident dependence of the
speeds on the source locations. The speed histograms of limb,
on-disk and all CMEs are quite similar. All of them
have the same peak around 200 -- 400 km s$^{-1}$ with the same average value
of about 435 km s$^{-1}$.
If the projection effect is taken into account, the on-disk CMEs
should be generally faster than limb CMEs. This result is
contrary to the study by \citet{Burkepile_etal_2004}, who
investigated 111 limb CMEs observed by SMM and found that their
average apparent speed is 519 km s$^{-1}$, significantly larger than
that of all SMM CMEs. They believe that the projection effect causes
limb CMEs to have a greater apparent speed than other CMEs. However,
according to our statistical result, we think that the selection
bias in their study rather than the projection effect might be the real
reason. The limb CMEs identified by them must be associated with a
clear eruptive prominence or X-ray/H$\alpha$ flare. The imposed
criteria possibly made them filter out many weak/slow limb
CMEs.\label{pg_selectionbias} This may also be the reason why their average speed
of limb CMEs is larger than ours.

The distribution of angular width for all, limb and on-disk CMEs are shown
in Figure~\ref{fg_speed}(c) and \ref{fg_speed}(d). Although the scattering plot
does not manifest any evident correlation between the width and position,
the histograms for limb and on-disk CMEs are different. The average
value of angular width is about 59$^\circ$ for limb CMEs, but 120$^\circ$ for
on-disk CMEs. Moreover, almost all ($\sim95\%$) of the full halo CMEs are on-disk CMEs.
It suggests that the projection effect is significant for on-disk CMEs. A
further discussion of the projection effect will be given in Sec.\ref{sec_halo}.
As the projection effect is minimized for limb CMEs, thus we think
that the width distribution for limb CMEs obtained here reflects the truth. About
65\% of limb CMEs have an angular width in the range of $30^\circ-90^\circ$. The
average angular width ($\sim59^\circ$) of the limb CMEs is consistent with that
($\sim47^\circ-61^\circ$) obtained in previous work
\citep[e.g.,][]{Burkepile_etal_2004, Yashiro_etal_2004}.

\section{Inferences and Implications}\label{sec_implications}
The information of CMEs' source locations allows us to perform a deeper analysis than before.
As has been mentioned in the section of Introduction, the following four issues will be
addressed.

\subsection{Missing Rate of CMEs}
\subsubsection{Visibility in EIT}\label{sec_eit}
First of all, it is known that not all of white-light CMEs could be seen in
LASCO cameras. How many CMEs will be missed? Before answering this question,
we will discuss the visibility of CMEs in EIT instrument.
A primary function of EIT in CME study is to learn the eruptive processes
of CMEs in corona. It is also a necessary tool to distinguish if a CME originates
from front-side or back-side solar disk. This becomes even important when
someone wants to predict the geoeffectiveness of CMEs.

The comparison of the percentages of LI, AL and NS CMEs (Table~\ref{tb_classification}),
we found that there are a significant fraction of CMEs probably missed by EIT 195 \AA\ wavelength.
It has been mentioned in Sec.~\ref{sec_data} that
AL CMEs are probably from backside disk but close to the limb. If all the NS CMEs are
considered as backside events, the percentage of backside CMEs (NS $+$ AL CMEs)
would be about 66\%, larger than the expected value 50\%. It implies that a
significant fraction, $\sim16\%$, of LASCO CMEs probably occurred on the
front-side solar disk, but did not leave any visible eruptive signatures in EIT
195 \AA\ images.

Further, the speed histograms in Figure~\ref{fg_speed}(b) suggests that there
is a jump around 200 km s$^{-1}$. The CMEs with speed less than 200 km
s$^{-1}$ occupy about ten percent only, and particularly, there is only
one CME slower than 100 km s$^{-1}$. However, the statistical analysis of all
LASCO CMEs did not show such a speed cut-off at low value (refer to Figure 4 in
\citealt{StCyr_etal_2000} and Figure 5 in \citealt{Yashiro_etal_2004}). Consider that our sample
includes LI CMEs only, thus the low rate of slow CMEs we obtained here
probably reflects the fact that there is a threshold of speed
somewhere between 100 and 200 km s$^{-1}$, and a CME with a speed less
than the threshold is generally too weak to leave an evident eruptive
signature on the solar surface.

\subsubsection{Visibility in LASCO}\label{sec_lasco}
To address the visibility of CMEs in LASCO,
we have to make two assumptions. The first one is that the limb CMEs are all visible to LASCO.
This assumption is reasonable because limb CMEs are supposed to be least affected by projection effect,
occulting effect and Thomson scattering effect. Also this assumption seems to
be true according to \citet{Yashiro_etal_2005} work.
The second assumption is that all the 16\% front-side CMEs missed by EIT have the
source location distribution as same as the LI CMEs. Then the above
question may be answered by studying the $DSC$ distribution of LI CMEs shown in Figure~\ref{fg_dsc}.

\begin{figure}[tb]
  \centering
  \includegraphics[width=0.8\hsize]{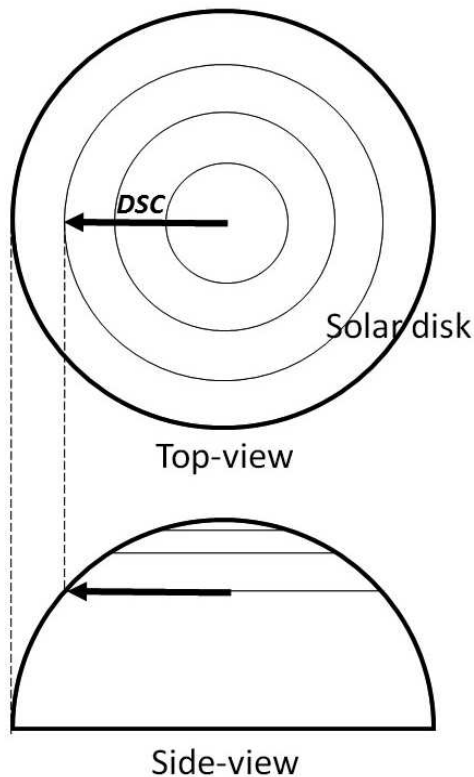}
  \caption{A schematic picture illustrates the effect of the spherical geometry on the CME
distribution with $DSC$.}\label{fg_sphere}
\end{figure}

If CMEs uniform-randomly occur on the
solar surface and all of them can be detected, the spherical
geometry will cause a non-uniform distribution of the CME occurrence rate with respect to $DSC$,
as illustrated by Figure~\ref{fg_sphere}. It is easily obtained that
the contribution of the spherical geometry to the probability distribution of
the CME counts is given by
\begin{eqnarray}
P=\left\{
\begin{array}{ll}
\multicolumn{2}{l}{
A\frac{2\pi
DSC}{\sqrt{1-DSC^2}}\left[1-\frac{2}{\pi}\arccos\left(\frac{\sin60^\circ}{DSC}\right)\right], }\\
&DSC>\sin60^\circ \\
A\frac{2\pi
DSC}{\sqrt{1-DSC^2}},\ \ \ \ \ \ \ \ \ \ \ \ \ \ \ \  &DSC\leq\sin60^\circ
\end{array}
\right.\label{eq_2}
\end{eqnarray}
in which the coefficient $A\approx0.2$ makes the integration of $P$ over $DSC$ is unity.
Here $\sin60^\circ$ corresponds to the latitude
of $\pm60^\circ$. This threshold must be set because there are few CMEs originating from the high
latitude regions (refer to Fig.\ref{fg_solardisk}).
Under the assumption that all limb CMEs are
visible to LASCO, the expected CME counts in each $DSC$ bin can be
calculated by Eq.\ref{eq_2}, which has been presented by the red histogram in Figure~\ref{fg_dsc}.

Comparing the red and black histograms,
we are able to estimate the missing rate of CMEs for SOHO/LASCO, which is given by the following formula
\begin{eqnarray}
\begin{array}{l}
\mathrm{Missing\ Rate} \\
=\frac{\mathrm{Expected\ CME\ Counts}-\mathrm{Recorded\ CME\ Counts}}{\mathrm{Expected\ CME\ Counts}} 
\end{array}
\end{eqnarray}
The blue symbols in Figure~\ref{fg_dsc} indicate the missing rate.
It is found that the missing rate roughly decreases with increasing $DSC$,
and on average, about 19\% of CMEs are not detected by the coronagraph.
This missing rate is slightly larger than that
obtained by \citet{Yashiro_etal_2005}, who investigated the CME
association of X-ray flares greater then C3 and found a missing rate of $\sim14\%$ averagely.
In their statistics, the missing rate increases as the associated flare intensity
decreases. Thus, their missing rate should be slightly underestimated,
because they did not consider flares weaker than C3.

\subsubsection{Invisible Front-side CMEs}\label{sec_invisible}
Combining the missing rates of white-light CMEs in LASCO and front-side
CMEs in EIT, we may infer that about 32\% of front-side CMEs can not be
recognized by SOHO. Recently, a concept of `stealth' CMEs has been proposed
to describe a kind of CMEs that do not leave any eruptive signatures in
EUV passbands and sometimes may not even be visible in coronagraphs facing
on them. The observations from STEREO twin spacecraft did support the existence of
such cases, like the 2008 June 1 event \citep{Robbrecht_etal_2009}.
This event totally had no eruptive signature in STEREO/EUVI images and was
extremely faint in the coronagraphs on board STEREO-B. If there was no
STEREO-A spacecraft, in which the CME was a limb event, it would probably be missed.
Here we call these 32\% of front-side CMEs `SOHO stealth' CMEs.

Frankly, the rate sounds too high, because there are only several cases
found in STEREO data. This is probably caused by some technique limits of SOHO.
For example, the cadence of SOHO/EIT is 12 minutes, and
therefore, a quick eruption lasting less than 12 minutes is
possible to be missed by EIT. Besides, low signal-to-noise ratio is
another possible technique reason. Of course, there is a physical explanation
that such a stealth CME might launch from an altitude not corresponding to the
designed EUV passbands of instruments, so that no signature can be observed.
No matter which reason is, it is clear that these stealth CMEs should be weak and probably travel
across a relatively small region on the solar surface.

The 32\% SOHO stealth CMEs provide us a reasonable explanation of the high rate
of the missing alert of geomagnetic storms and/or ICMEs encountering the Earth.
The association of ICMEs with CMEs has been studied by many researchers before
\citep[e.g.,][]{Lindsay_etal_1999, Cane_etal_2000, Gopalswamy_etal_2000,
Cane_Richardson_2003}, and was summarized in the review by
\citet{Yermolaev_Yermolaev_2006}. A fact revealed by these investigations is that
the association rate is not 100\%, and about 18 -- 44\% of ICMEs can not be found
the corresponding CMEs. The missing rate obtained from our study is in highly
agreement with these previous results.

A direct consequence is that there would be a significant fraction of geomagnetic
storms are probably not able to be predicted. \citet{Webb_etal_1998} in their
paper discussed so-called first `problem' storm, the 1997 January 17 event.
This event was thought no CME was observed to associate with, though the interplanetary
shock ejecta pair causing this storm was obvious. There are many other problem storms,
which can be found in the studies by, e.g., \citet{Webb_etal_2000},
\citet{Schwenn_etal_2005}, and \citet{Zhang_etal_2007}. The existence of
stealth CMEs is a natural explanation of such storms. There are also other
explanations. One is the longitudinal extension of CMEs \citep{Webb_etal_2000, Zhang_etal_2003},
and the other is the CME deflections \citep{Wang_etal_2004b, Wang_etal_2006a}.
The latter will be discussed further in Sec.\ref{sec_deflection}.

\subsection{Mass of CMEs}\label{sec_dis2}
The previous studies suggested that the mass of a typical CME is about
$10^{12}$ kg \citep[e.g.,][]{Vourlidas_etal_2000}. The value
is estimated according to the brightness of the transient structure detected
in coronagraphs. Actually the enhanced brightness is not only contributed by
the CME, but also the compressed solar wind plasma surrounding the CME. Before discussing
this issue, we must clarify the definition of CME. The CME was first observed
by the white-light coronagraph onboard OSO-7 in December
1971, and defined as an enhanced bright erupting structure, i.e., the luminescence area in the FOV
of a coronagraph. However, this
graphic definition is not strict, because ambient solar wind plasma may be disrupted
by a CME and caused brightened. Thus, here a CME strictly refers to
the plasma ejected away from the solar atmosphere that does not include the
disrupted solar wind. As to the luminescence area in the FOV of a coronagraph
during a CME, we call it transient. What we will study below is whether or not
the transient contains only the CME.

\begin{figure}[tb]
  \centering
  \includegraphics[width=\hsize]{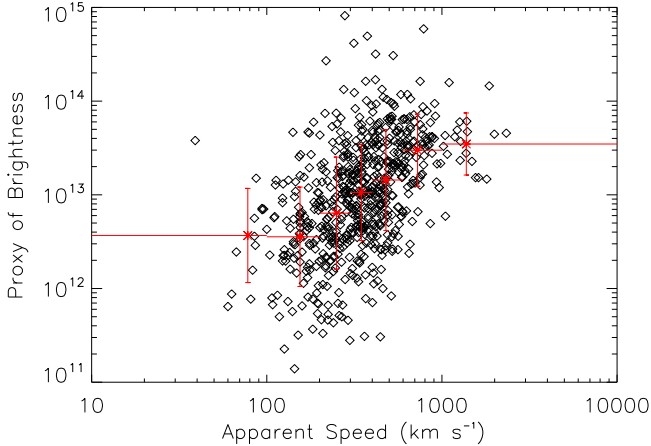}
  \caption{The scattering plot of the CME brightness proxy versus the apparent speed for
the CMEs from 1997 to 1998. The red symbols with error bars mark the average
value of the brightness proxy within the speed range indicated by the associated
horizontal red lines.}\label{fg_brightness}
\end{figure}

There is the parameter of mass listed in CDAW CME catalog. The mass is estimated by the method
developed by \citet{Vourlidas_etal_2000} based on observations of
white-light coronagraphs and some assumptions. Briefly, the estimated mass
of a transient is the product of the size of the luminescence region and the
value of the enhanced brightness. Angular width
is an important parameter to weight the size of the region though we do not have
the information of the span of the region along the radial direction. Thus,
as a first-order approximation, we may derive a new parameter, which is the ratio of
the mass to the angular width, and treat it as a proxy of the brightness of a transient.
Figure~\ref{fg_brightness} presents the brightness proxy versus
the apparent speed, in which all the CMEs listed in the CDAW CME catalog
during 1997 -- 1998 are included. Surprisingly, there is an
obvious positive correlation between the two parameters. Actually, this
phenomenon has been implied in Figure~\ref{fg_speed}(b), which shows that
the distributions of speeds for limb and on-disk CMEs are quite similar.
It should be pointed, however, that the correlation coefficient is only
0.5 and the scattering is significant. We think that such large scattering may be
due to the inaccurately estimation of the mass, angular width, etc. Overall, there is a trend that
transients with a slower apparent speed look fainter in a coronagraph.

Let's compare
two identical transients with the same apparent speed. Transient-1 rises above the
limb and Transient-2 comes from the longitude of 30$^\circ$ and equator. The two transients should have the same
brightness according to the above analysis, but the real speed and heliocentric distance
of Transient-2 should be twice as large as those of Transient-1 if the projection effect is taken into account.
So tracking Transient-2 back to the heliocentric distance of Transient-1, its brightness should be doubled.
By assuming that the speed of the transient changes little within that region, it can be inferred
that the brightness of a transient at any given heliocentric distance should be positively correlated
with its real speed.

Why is the brightness of a transient controlled by its speed? It can be
easily explained if a transient contains not only a CME but also the ambient compressed
solar wind plasma due to the CME. The faster a CME is, the greater is
compression of the ambient solar wind plasma, and therefore the brighter the transient looks.
This picture confirms and deepens the previous thought of three-component-structure CMEs that the
bright fronts of CMEs are believed to be the compressed solar wind. Our result obtained
here is suitable for any type CMEs. Thus, the mass given in the CDAW CME catalog is not
merely the CME mass, but the mass of both the CME and the ambient compressed solar wind
plasma, which we can call it `apparent mass'.

\begin{figure*}[tb]
  \centering
  \includegraphics[width=0.32\hsize]{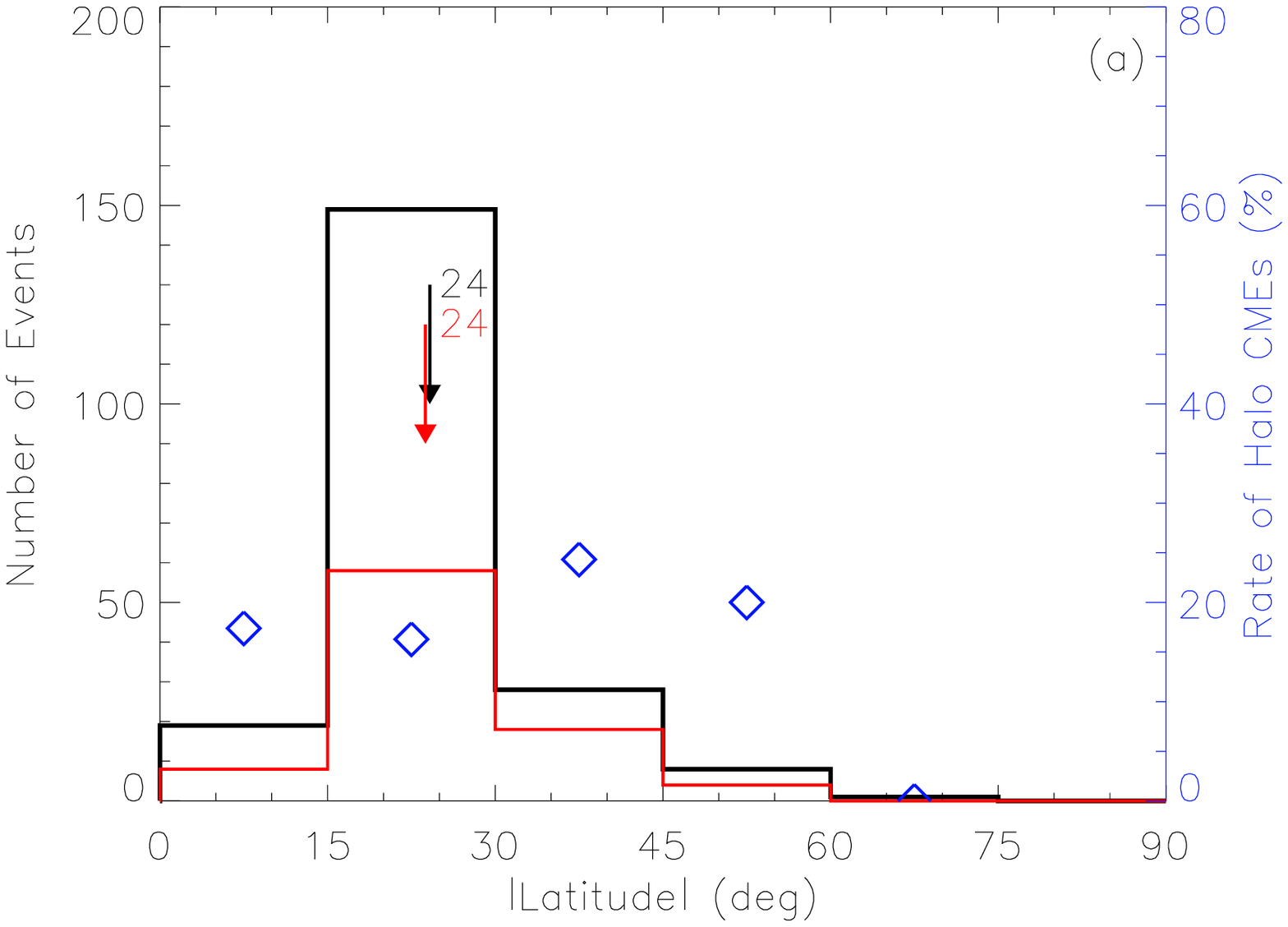}
  \includegraphics[width=0.32\hsize]{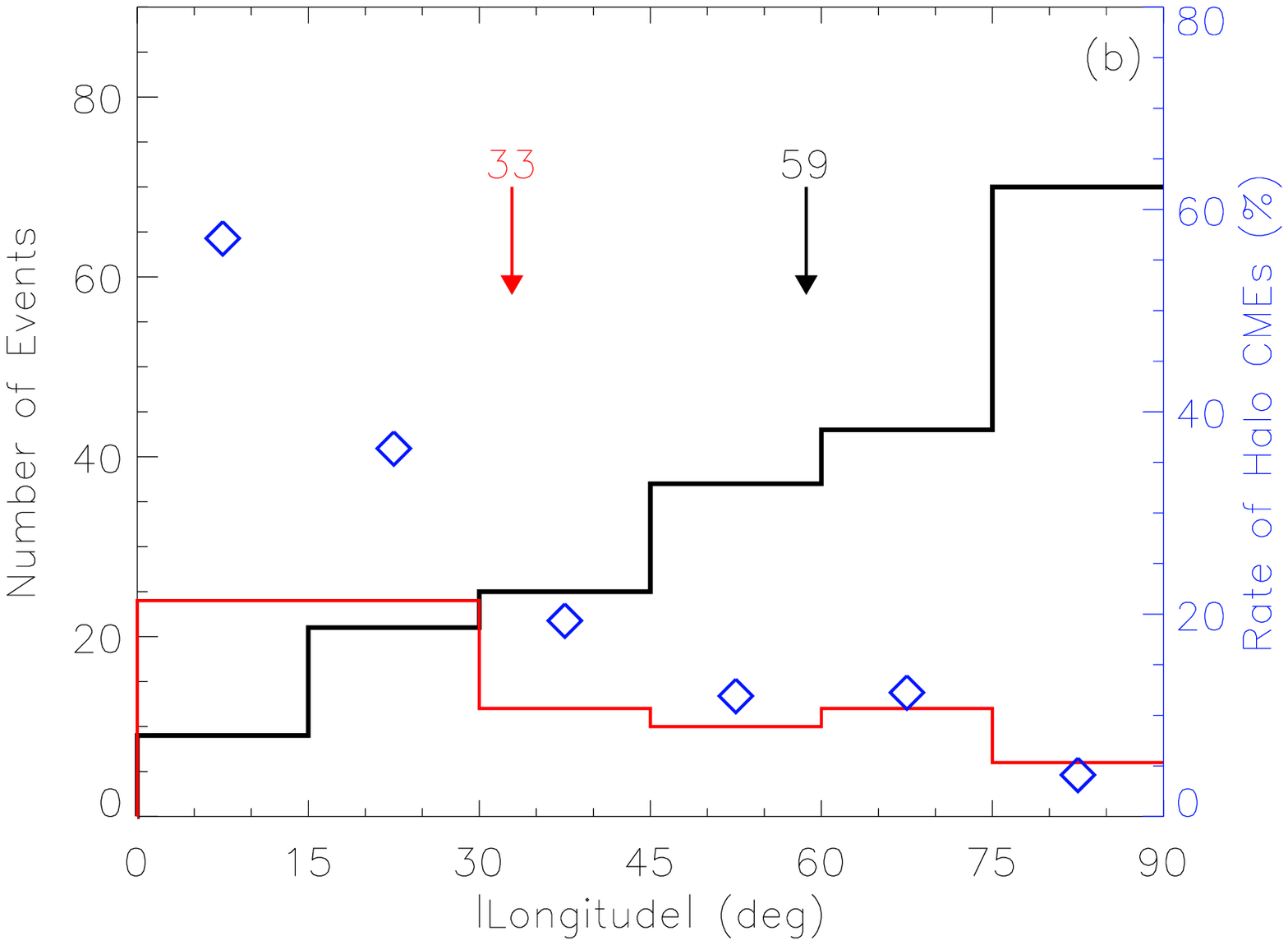}
  \includegraphics[width=0.32\hsize]{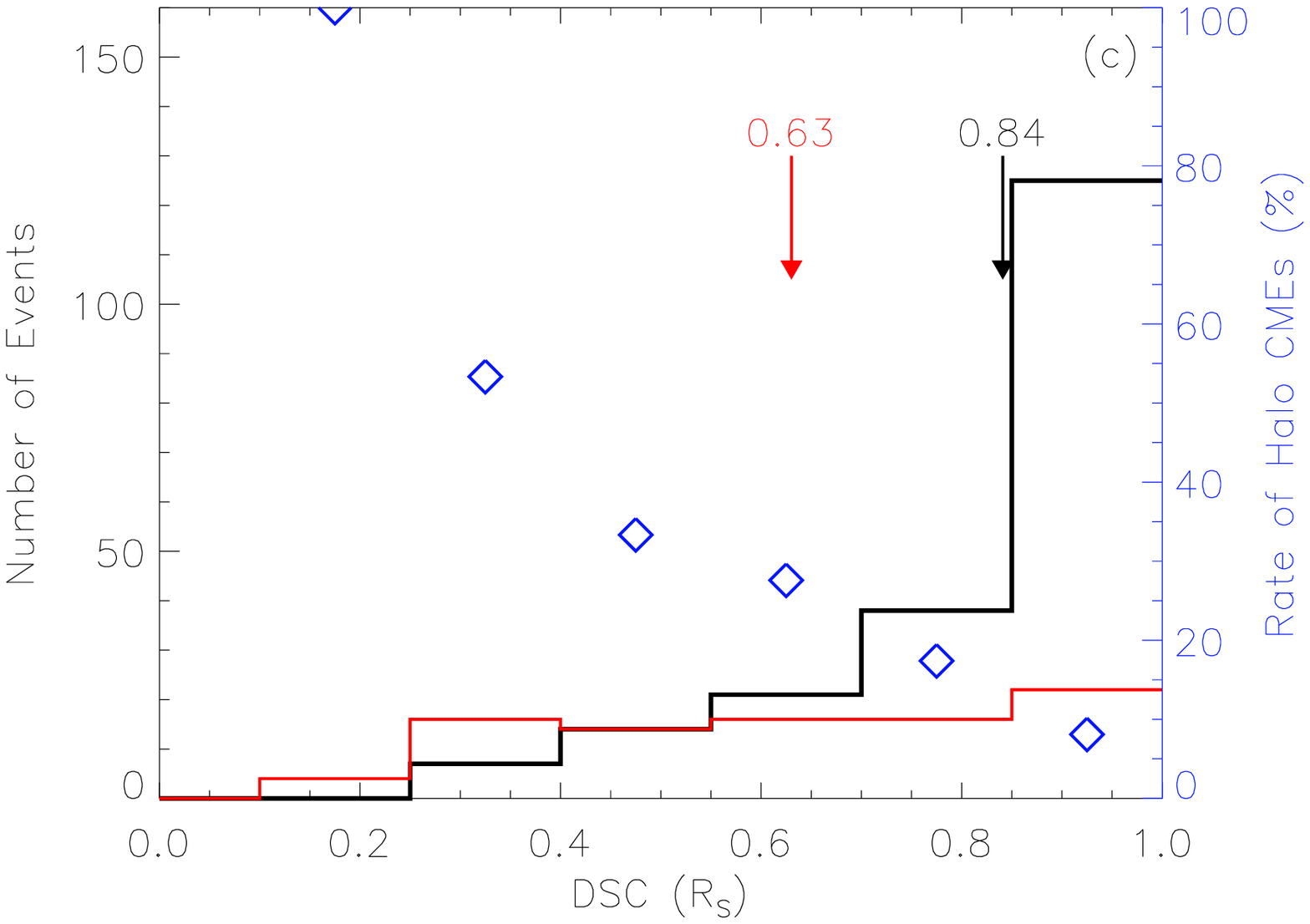}
  \caption{The histograms of absolute value of latitude (panel a), absolute value of longitude (panel b)
and $DSC$ (panel c) for halo (red) and non-halo CMEs (black), respectively. The blue diamond symbols
denote the rate of halo CMEs. For clarity, the counts of halo CMEs are multiplied by a factor of 2.}\label{fg_halo_dsl}
\end{figure*}
\begin{figure*}[tb]
  \centering
  \includegraphics[width=0.32\hsize]{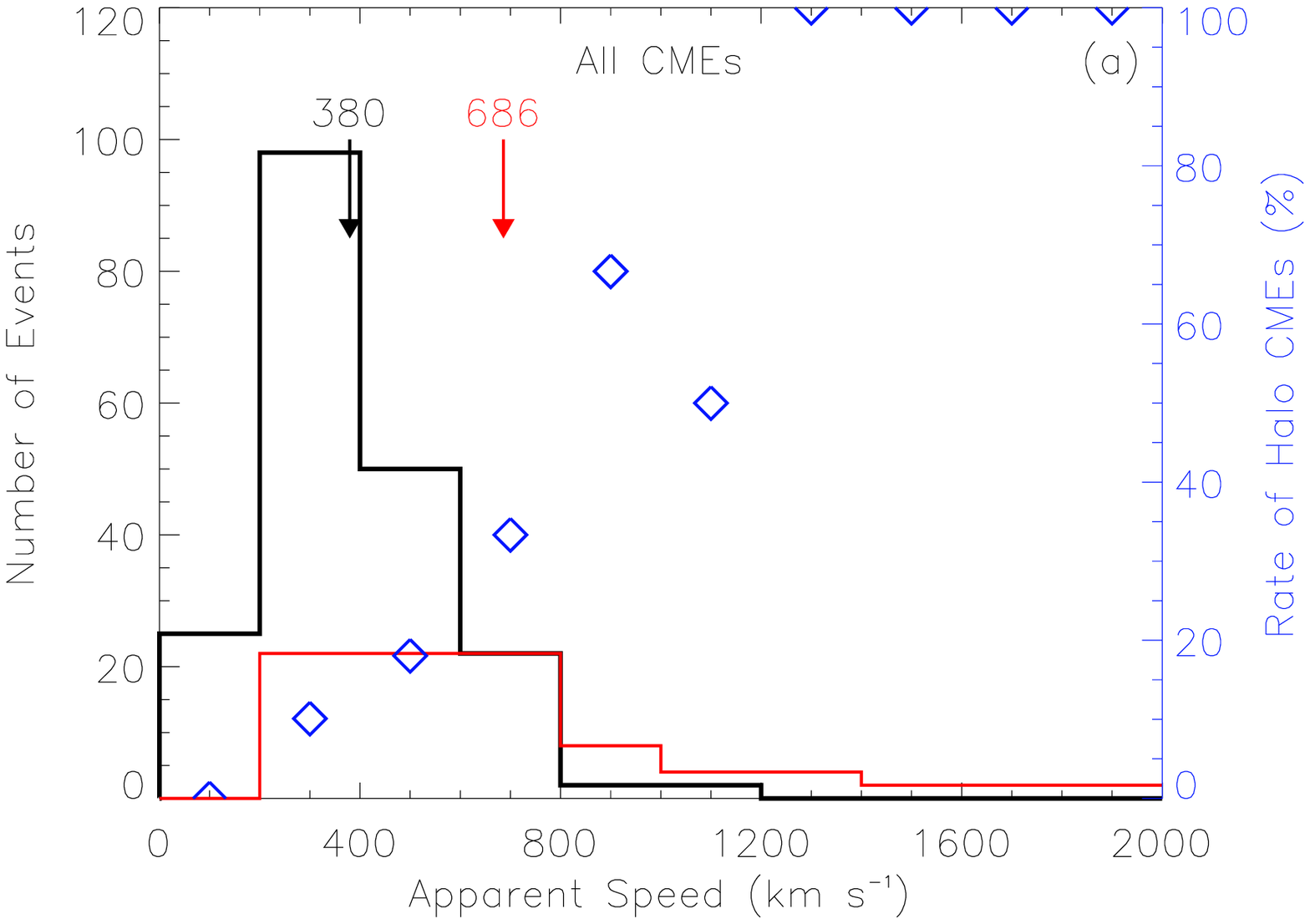}
  \includegraphics[width=0.32\hsize]{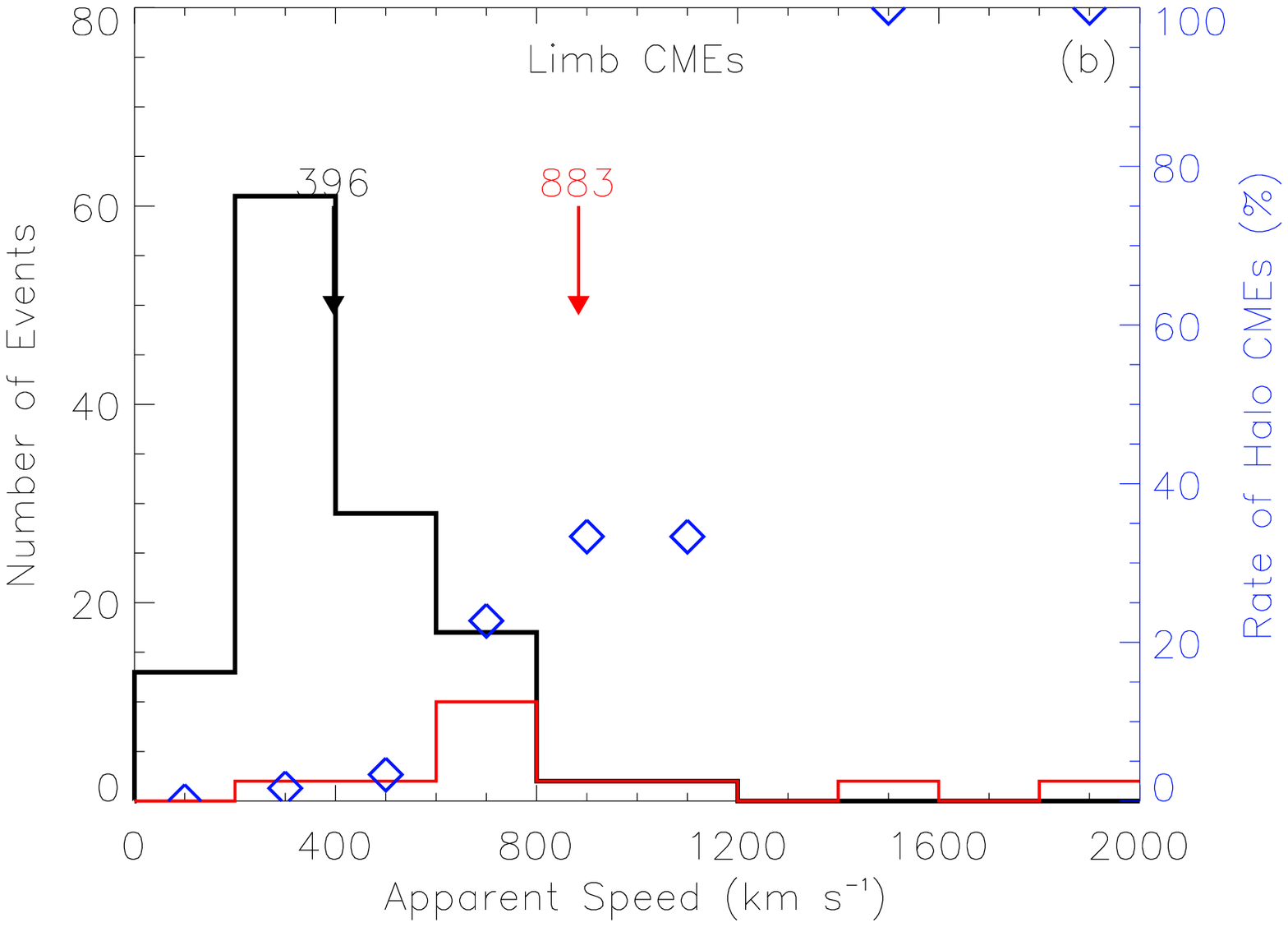}
  \includegraphics[width=0.32\hsize]{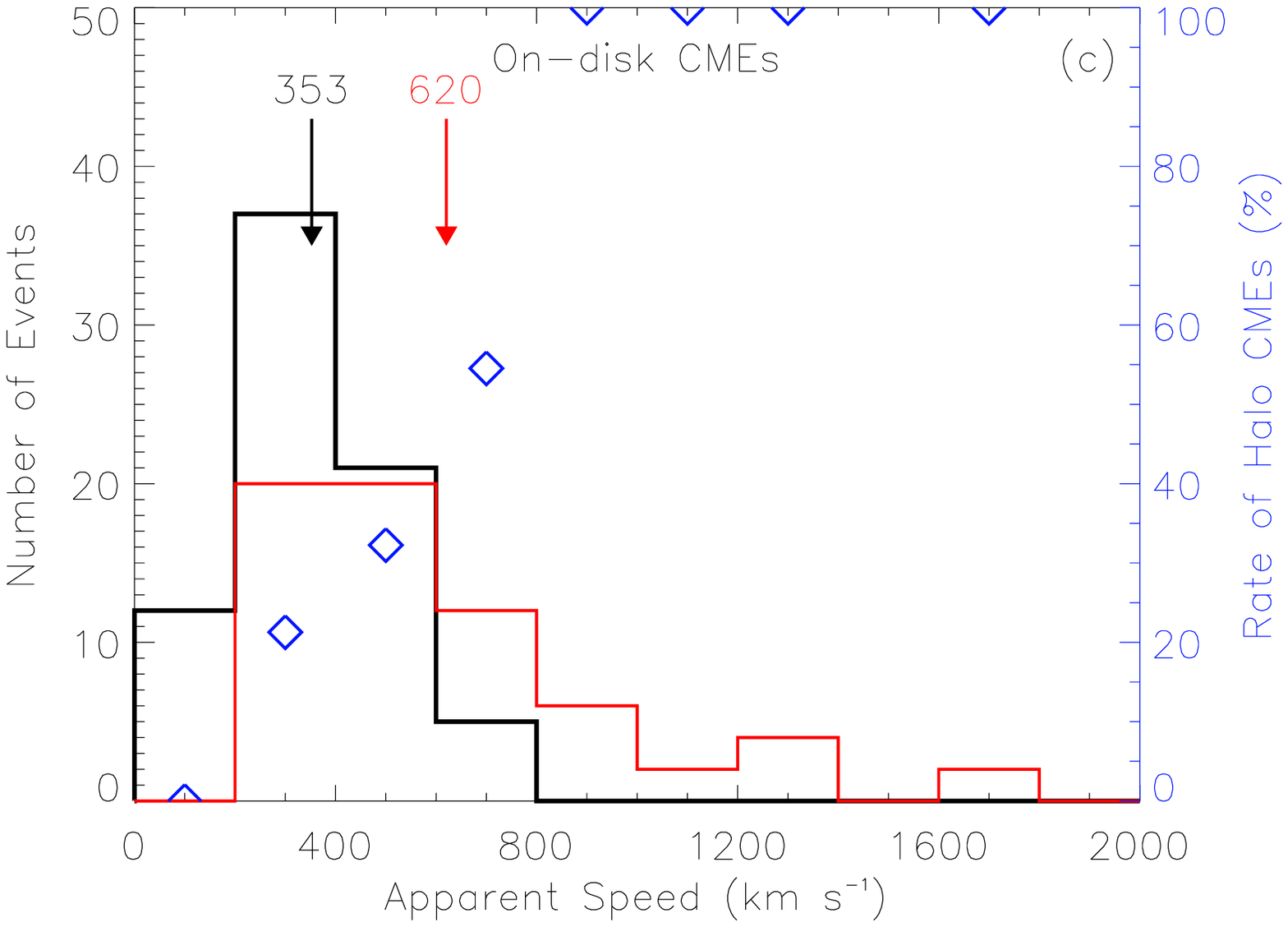}
  \caption{From (a) to (c), the panels show the histograms of apparent speed for all, limb and on-disk
CMEs, respectively. The pattern of them is as the same as that in Figure~\ref{fg_halo_dsl}.}\label{fg_halo_speed}
\end{figure*}

To a certain extent, the mass component contributed by the compressed solar wind plasma
stands for the `virtual' mass, which is a concept first
proposed in fluid mechanics. Briefly, the presence of virtual mass is because `an
accelerating or decelerating body must move some volume of surrounding fluid as it
moves through it, since the object and fluid cannot occupy the same physical space
simultaneously\footnote{Adapted from Wikipedia \url{http://en.wikipedia.org/wiki/Added_mass}}.'
Since CMEs propagate in solar wind, the concept of virtual mass is also applicable to the
CME studies. In practice, the apparent mass is obtainable, but not the CME mass or
virtual mass, and scientists are used to using the apparent mass as the CME mass.
In that situation, the CME mass is overestimated, and the CME volume is obviously
also overestimated.
These overestimations will lead to uncertainties or errors in other relevant CME studies,
e.g., the CME trigger and initiation, the CME aerodynamics in IP space.
If we can not accurately estimate the kinetic energy of a CME, we may not really
understand how a CME energy is accumulated and released. If we do not know the
volume and mass of a CME precisely, we may add a wrong virtual mass in the governing
equation of CME aerodynamics and also may be difficult to figure out how large the drag force
acting on a CME is when it propagates in corona and interplanetary space (previous researches
can be found in, e.g., \citealt{Cargill_etal_1996, Cargill_etal_2004}).
Some deeper discussions about these issues are worthy of being pursued in another paper.

\subsection{Causes of Halo CMEs}\label{sec_halo}
Halo CMEs generally get special attention of many researchers as
they have a higher probability to hit the Earth. This is because
people believe that the projection effect is the main cause of
a CME looking halo. However, it is not the only cause. This point
can be seen by comparing the distributions of halo and non-halo CMEs.

\subsubsection{Halo vs. Non-halo CMEs}
As has been defined before, we consider a CME to be halo when its
apparent angular width is larger than $100^\circ$, otherwise the CME is
non-halo one. The numbers of halo and non-halo CMEs have been listed
in Table~\ref{tb_sources}.
The similarity and difference between the two kinds
of CMEs in the distribution of source locations are given in
Figure~\ref{fg_halo_dsl}. The latitude distributions of halo and
non-halo CMEs are similar to each other (Fig.\ref{fg_halo_dsl}(a)), and both of them have a
peak at around $\pm(15^\circ-30^\circ)$. The rate of halo CMEs, i.e., the ratio of
the number of halo CMEs to the number of all CMEs (indicated
by the blue symbols), is around
20\%. The longitude
distributions of them are quite different (Fig.\ref{fg_halo_dsl}(b)). For halo CMEs,
the average longitude is about $33^\circ$, and the CME count
decreases as the absolute value of longitude
increases; whereas for
non-halo CMEs, the average longitude is about $59^\circ$, and the count
increases. It could be found that the longitude distribution of halo
CMEs is relatively flat comparing to that of non-halo CMEs. The rate of halo CMEs
has a clear decrease trend from central meridian to limb. Within $\pm15^\circ$, the
rate reaches $\sim57\%$, while outside of $\pm75^\circ$ the rate is as low as $\sim4\%$.

As did before, we further investigate the parameter $DSC$, which is
shown in Figure~\ref{fg_halo_dsl}(c). The distributions of the two
kinds of CMEs present a substantial difference. Most non-halo
CMEs come from the regions far away from the disk center. It can be
estimated that there are about 62\% of non-halo CMEs with $DSC$ equal or larger
than 0.85 $R_S$, and
no non-halo CME originating from central region with $DSC$
less than 0.25 $R_S$. On contrary, the distribution of halo CMEs is
much flatter, indicating that halo CMEs could originate anywhere. The rate of halo CMEs monotonically
decreases from 100\% to $\sim8\%$ with the increasing $DSC$. However, one may notice that the highest
peak of the distribution of halo CMEs appears in the $DSC$ range of 0.85 -- 1.0
$R_S$, which occupies 25\% of the halo CMEs. These results imply
that (1) projection effect is indeed one factor causing a CME looking halo, but
(2) it is not the only one factor, especially for those halo CMEs close to the solar limb. As will
be seen below, the `violent eruption' probably is the other
major cause. The term `violent eruption' here means an eruptive process, during which the
released energy is higher than usual and the release process is quicker.

\citet{Lara_etal_2006} had exclusively addressed the issue whether
or not halo CMEs are special events, and reached the conclusion that
the behavior of halo CMEs can not explained merely by projection
effect. They believe that `the observed halo is the manifestation
of the shock wave driven by fast CMEs'. To drive a shock wave, the CME
speed must be larger than local Alfv\'{e}nic speed, i.e., halo CMEs should
be much energetic. Thus, their interpretation on halo CMEs
is more or less consistent with our second point of view given in the last paragraph.
Moreover, that point can be further clarified by the comparison of
the speed distributions among the halo/non-halo and on-disk/limb CMEs.

\subsubsection{Limb vs. On-disk CMEs}
Figure~\ref{fg_halo_speed}(a) suggests that the average speed of halo CMEs is
generally twice as large as that of non-halo CMEs.
Moreover, for CMEs with speed $\leq 800$ km s$^{-1}$, the rate of halo CMEs is about 14\%, while, for
CMEs with speed $> 800$ km s$^{-1}$, the rate jumps to $\sim73\%$, and
these fast halo CMEs occupies about 25\% of all halo ones. Particularly, all the five CMEs
with speed larger than 1200 km s$^{-1}$ are halo. Since there are few non-halo CMEs
faster than 800 km s$^{-1}$, the percentage 25\% roughly indicates how many halo CMEs are faster,
i.e., more energetic, than the average level of CMEs.

To reduce the projection effect in our analysis, we investigate the
limb and on-disk CMEs separately. Figure~\ref{fg_halo_speed}(b) is for limb CMEs,
in which the halo CMEs occupy a percentage of $\sim8\%$. It is evident that
the distributions of halo and non-halo CMEs are much different. The peak of the distribution of
halo CMEs locates between 600 -- 800 km s$^{-1}$ with the average
value of about 883 km s$^{-1}$, and about $36\%$ of halo CMEs have an
apparent speed larger than 800 km s$^{-1}$. On the contrary, the
peak of non-halo CMEs is between 200 -- 400 km s$^{-1}$ with the
average value of 396 km s$^{-1}$, and only about $3\%$ of them are
faster than 800 km s$^{-1}$. Thus, we think that the violent eruption
is the dominant reason for a limb CME to be halo.

For on-disk CMEs (Fig.\ref{fg_halo_speed}(c)), in which there are about 30\% halo CMEs,
the average speed of halo CMEs (620 km s$^{-1}$) is also
nearly twice of that of non-halo CMEs (353 km s$^{-1}$).
All the on-disk CMEs with an apparent speed larger than 800
km s$^{-1}$ are halo ones, which occupies about 21\% of the entire
on-disk halo CMEs. These results suggest that the violent eruption is at least one of the major causes
of halo CMEs. On the other hand, comparing Figure~\ref{fg_halo_speed}(b) and \ref{fg_halo_speed}(c),
we find that the histograms of on-disk and limb non-halo CMEs are similar, while
the histogram of on-disk halo CMEs is quite different from that of limb halo CMEs.
For limb CMEs, most halo CMEs are faster than 600 km s$^{-1}$. However, for on-disk CMEs,
most halo CMEs are slower than 600 km s$^{-1}$, which indicates that projection effect
is still a non-ignorable factor leading an on-disk CME to have a halo appearance.

In summary, both projection effect and violent eruption are the major causes
of halo CMEs. The projection effect being a cause is because (1)
the rate of halo CMEs monotonically decreases from 100\% to 8\% as
the CME source location moves from disk center to limb, (2)
no non-halo CMEs originated from the regions with $DSC<0.25 R_S$, and
(3) the average angular width of on-disk CMEs is $\sim120^\circ$,
twice of that of limb CMEs, and about 95\% of full halo CMEs are on-disk CMEs.
The second cause can be seen from the following facts. (1) About 25\% of
halo CMEs originating from solar limb ($DSC\geq0.85R_S$), where the projection
effect is minimized. (2) The apparent speed of halo CMEs is averagely
twice of that of non-halo CMEs, no matter whether they are on-disk or limb
CMEs. (3) Most of fast CMEs are halo CMEs; especially the rate is 100\% for speed
$> 1200$ km s$^{-1}$. Besides, for limb halo CMEs, the violent eruption
is the primary cause. Overall, there are about 25\% of halo CMEs above the
average level of CME energy.

\subsection{Deflections of CMEs}\label{sec_deflection}

\begin{figure}[tb]
  \centering
  \includegraphics[width=\hsize]{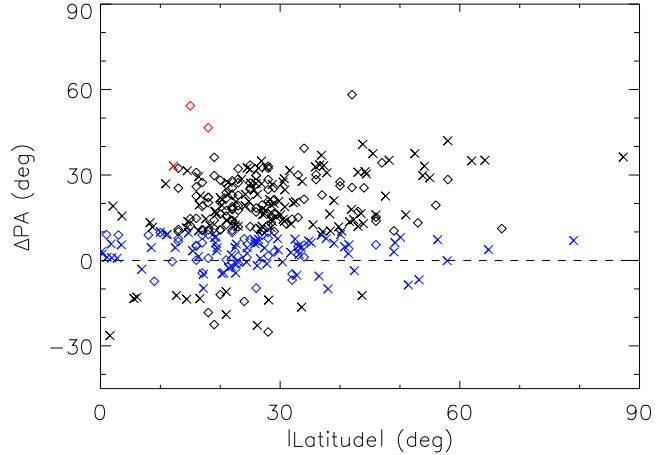}
  \caption{A scattering plot showing the deflection angle as a function
of the absolute latitude of the CME source location. A data point above the dashed
line at zero means the CME deflected towards equator, otherwise towards
polar regions. The blue symbols mark the events with deflection angle
less than $10^\circ$, which we treated as radial events. The red symbols
mark the events, whose deflection crosses over the
equator by more than $20^\circ$.}\label{fg_def0}
\end{figure}

\begin{figure*}[tbh]
  \centering
  \includegraphics[width=0.8\hsize]{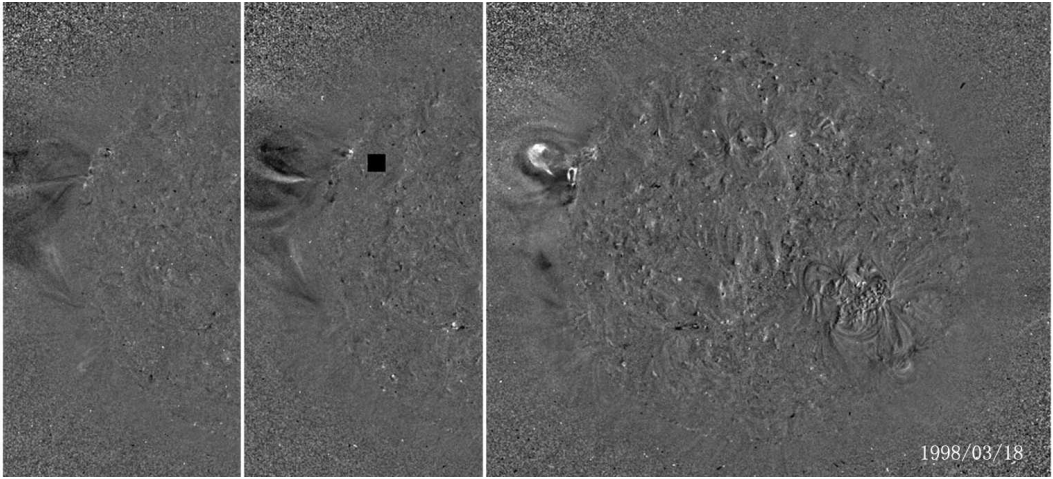}
  \caption{Asymmetrical expansion of a CME on 1998 March 18, which resulted in the significant
difference between $SPA$ and $CPA$.}\label{fg_def1}
\end{figure*}
\begin{figure*}[tbh]
  \centering
  \includegraphics[width=0.8\hsize]{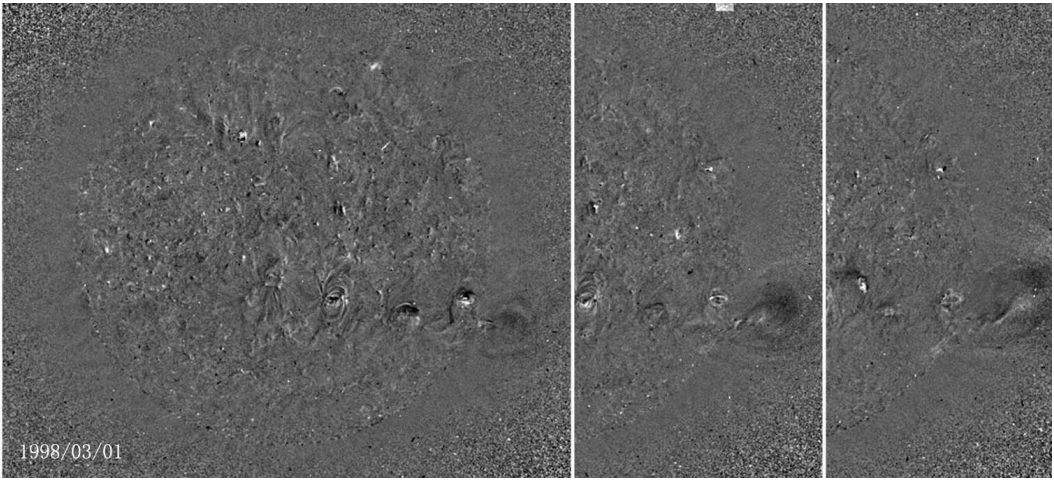}
  \caption{Non-radial ejection of a CME on 1998 March 1.}\label{fg_def2}
\end{figure*}
\begin{figure*}[tbh]
  \centering
  \includegraphics[width=0.8\hsize]{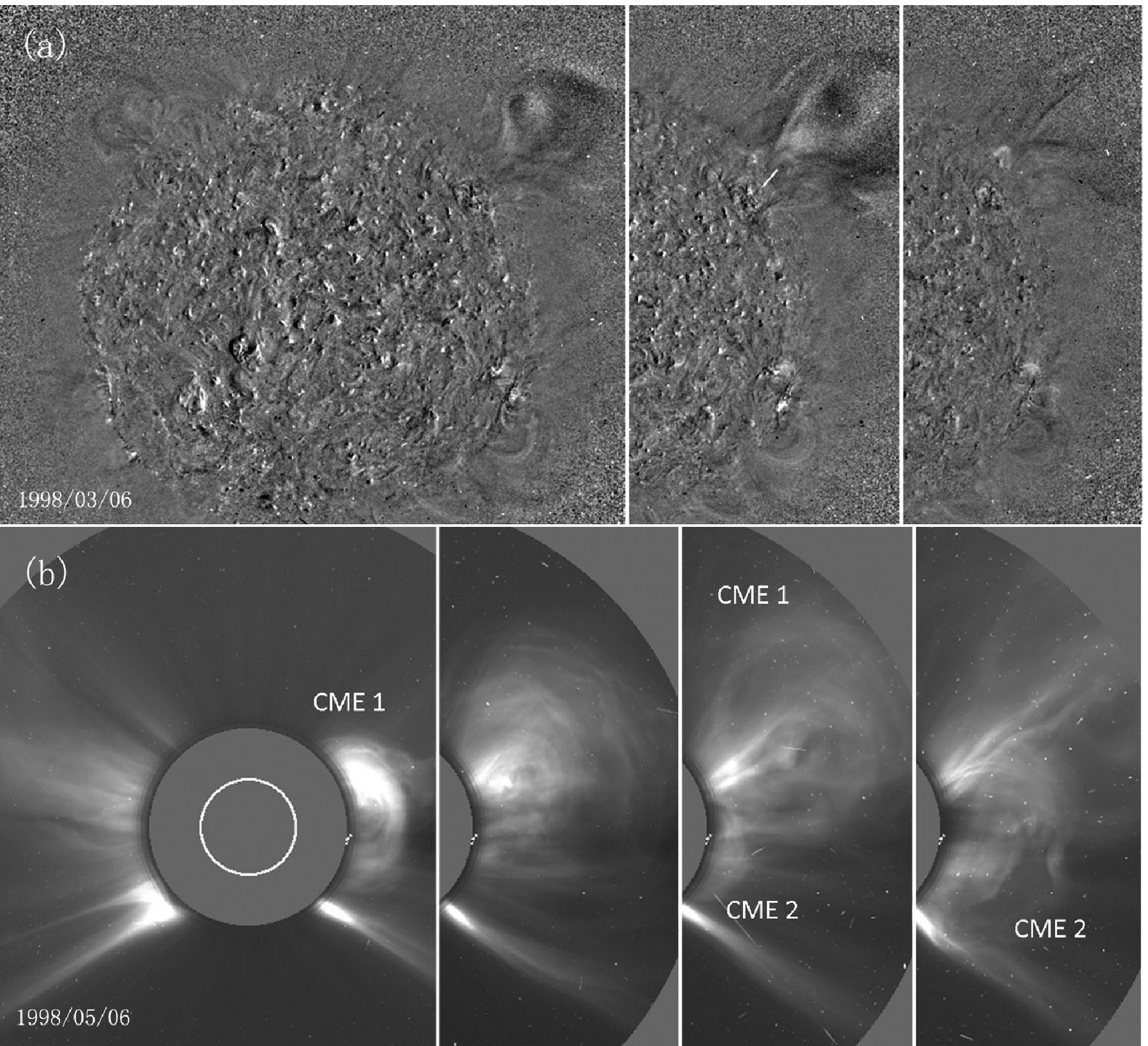}
  \caption{Panel (a): Curved propagation of a CME viewed in FOV of EIT on 1998 March 6. Panel (b):
Deflected propagation of a CME (CME1) viewed in FOV of LASCO on 1998 May 6, which is caused by
the collision of CME1 with a following CME (CME2).}\label{fg_def3}
\end{figure*}

\subsubsection{Statistical Results and Classification}\label{sec_dclass}
In our dataset, we have the parameter $SPA$ (measured in EIT images)
of LI and AL CMEs, and therefore the deflections of CMEs can be studied by
comparing $SPA$ with the central position angles ($CPA$, measured in LASCO) of CMEs.
To reduce the projection effect, only the 138 limb LI CMEs and 191 non-halo AL CMEs
with available position angles and $CL$ equal to 1 and 2 are considered. To obtain
the direction and magnitude of the deflection of these CMEs, we
calculate the difference ($\Delta PA$) between $SPA$ and $CPA$.
Figure~\ref{fg_def0} presents $\Delta PA$ as a function of
the absolute latitude of CME's source location.
LI CMEs are denoted by diamonds, and AL CMEs by `$\times$'.
Note that, for AL CMEs,
we use $SPA$ to estimate the approximate latitudes of their source locations, as
we believe that these CMEs occurred near the solar limb though they were on the backside.
A positive value of $\Delta PA$ means an equator-ward
deflection while a negative value corresponds to a pole-ward
deflection.
Considering the error in determining the CME's source location
and $CPA$, we treat the CMEs with $|\Delta PA|<10^\circ$
as radial events (the blue symbols in the figure).

From Figure~\ref{fg_def0}, it is obtained straightforwardly that (1) about 62\% of CMEs
underwent an equator-ward deflection with the average deflection angle of
$\sim22^\circ$, (2) about 5\% of CMEs manifested a significant pole-ward deflection
with the average angle of $\sim16^\circ$,
and (3) at high latitude regions (outside of $\pm45^\circ$), most (21 out of 31) CMEs deflected
towards equator and no CME towards the polar region.

Note that the deflections obtained here is simply from the comparison of the position
angle of the CME eruptive signature in EIT FOV and the CME central position angle in
LASCO FOV. Any measurement errors and inconsistency between the two measurements will
result in a faked
deflection. For many CMEs, the errors of the coordinates of their source locations are about 10 degrees,
which have been considered in the above analysis. However, CMEs are a large scale
structure. Their source regions may span over a large area, and the identified
source locations in EIT images may possibly be not centered beneath CMEs
\citep[e.g.,][]{Harrison_1995, Harrison_Lyons_2000, Plunkett_etal_2001}. A quick check of the EIT
movies, we find that some CMEs do have two widely separated footpoints, and the
identified source locations (i.e., the most significant eruptive features) are close
to one of them, e.g., the 1998 April 20 and 1998 December 7 CMEs. For such cases,
the derived $\Delta PA$ are probably not correct or suffer a much larger error.

Although the faked deflections do exist, most deflections of CMEs in our statistics
are true. We find that these deflection behaviors can be classified into three types.
The first type is the manifestation of \emph{asymmetrical expansion} of CMEs. As a case,
Figure~\ref{fg_def1} shows the EIT images of the CME on 1998 March 18. At the
beginning, there was a clear flux-rope structure standing right-up on the eastern
limb. As the eruption progressed, the CME flux rope expanded asymmetrically. Its
boundary close to the equator was freely expanding, but the expansion of the
boundary close to the pole was obviously blocked by something. This asymmetrical expansion
caused the CME deflecting to the equator.
The second type is the \emph{non-radial ejection}. The CME occurring on 1998 March 1 belongs
to this type, as shown in Figure~\ref{fg_def2}. The CME flux rope was inclined toward
equator before the eruption, i.e., the direction of its ejection is non-radial
initially. Such configuration of the CME naturally leads to a non-radial propagation after
the non-radial ejection. The last type is called \emph{deflected propagation}. Not like the
second type, this type of deflections is mainly due to the interaction of CMEs with
other neighboring structures during their propagation in the corona. The neighboring structures
could be either the magnetic fields from coronal holes or other CMEs. For example,
Figure~\ref{fg_def3}(a) presents a CME on the western limb on 1998 March 6, whose trajectory in the FOV
of EIT is curved. Apparently, the curved propagation of the CME is due to the presence
of the polar magnetic field. The deflections caused by the interaction between CMEs is
demonstrated in Figure~\ref{fg_def3}(b). The CME (labeled as CME1), to be deflected,
appeared above the western limb in the FOV of LASCO on 1998 May 6, and initially propagated along the
position angle of $\sim270^\circ$. However, its trajectory was quickly deflected toward
the north pole due the collision of the CME with a following CME (labeled as CME2).
The collision of CMEs causing CME deflections was reported by \citet{Gopalswamy_etal_2001c},
and also studied with numerical simulations by \citet{Xiong_etal_2006, Xiong_etal_2009}.

\begin{table}[tbh]
\renewcommand{\arraystretch}{1.8}
\begin{center}
\caption{Event List of Unusual Deflections} \label{tb_deflection}
\tabcolsep 1pt
\footnotesize 
\begin{tabular}{c|ccccccc}
\hline
No.& Date Time &Location &$SPA$ &$CPA$ &$\Delta PA$ &Width &Speed   \\
& UT& &deg &deg &deg &deg &km s$^{-1}$ \\
\hline
&\multicolumn{7}{c}{Pole-ward Deflection} \\
P1 & 1998/11/25 06:30:05  &  N18 E72  &   71    &  53   &  -18  &   41  &   256   \\
P2 & 1998/11/25 14:30:05  &  N20 E73  &   69    &  57   &  -12  &   52  &   213   \\
P3 & 1998/11/26 11:30:06  &  N19 E57  &   68    &  45   &  -23  &   50  &   216   \\
P4 & 1998/12/04 21:30:10  &  S24 W56  &   241   &  227  &  -14  &   65  &   238   \\
P5 & 1998/12/07 15:30:05  &  N28 W62  &   302   &  327  &  -25  &   42  &   490   \\
P6 & 1998/02/24 18:27:05  &  -        &   264   &  251  &  -13  &   43  &   259   \\
P7 & 1998/03/17 15:06:15  &  -        &   134   &  146  &  -12  &   6   &   204   \\
P8 & 1998/05/09 15:18:25  &  -        &   242   &  228  &  -14  &   46  &   533   \\
P9 & 1998/05/30 23:28:13  &  -        &   264   &  251  &  -13  &   63  &   594   \\
P10 & 1998/06/02 21:06:24  &  -        &   69    &  50   &  -19  &   61  &   782   \\
P11 & 1998/06/17 06:55:18  &  -        &   284   &  298  &  -14  &   23  &   632   \\
P12 & 1998/10/28 04:54:05  &  -        &   56    &  40   &  -16  &   59  &   208   \\
P13 & 1998/10/28 07:54:05  &  -        &   107   &  120  &  -13  &   66  &   486   \\
P14 & 1998/11/01 08:18:09  &  -        &   291   &  302  &  -11  &   25  &   238   \\
P15 & 1998/11/10 01:54:05  &  -        &   257   &  245  &  -12  &   29  &   284   \\
P16 & 1998/11/12 05:54:06  &  -        &   244   &  221  &  -23  &   19  &   254   \\
P17 & 1998/12/06 10:54:05  &  -        &   92    &  118  &  -26  &   73  &   806   \\
\hline
&\multicolumn{7}{c}{Equator-ward Overshooting} \\
O1 & 1998/03/18 07:33:06  &  N18 E87  &   71    &  118  &   47  &   174  &   636   \\
O2 & 1998/05/06 08:29:13  &  S15 W67  &   255   &  309  &   54  &   190  &   1099  \\
O3 & 1998/06/15 06:55:20  &  -        &   258   &  291  &   33  &   93  &   535   \\
\hline
\end{tabular}\\
The CMEs without identified locations are the AL events.
\end{center}
\end{table}

\subsubsection{Interpretation and Exceptions}
Except the deflections caused by interactions of CMEs, which is not frequent during
the solar minimum, almost all the CME deflections are essentially due to the pre-existing
magnetic field structures surrounding the CMEs.
For asymmetrical expansions and deflected propagations, the presence of the
ambient coronal magnetic field and solar wind originating from
the neighboring coronal holes may play a major role \citep[e.g.,][]{Gopalswamy_etal_2004},
while for non-radial ejections, the magnetic field configuration in
CME source regions decides the CME launch directions.
Near the solar minimum, the Sun has an approximately
dipole field, coronal holes usually appear in the polar regions,
from which the magnetic field and solar wind super-radially disperse
towards the equator, and therefore CMEs are deflected. This picture
is in agreement with the statistical result obtained here that most
CMEs propagated towards the equator. A model based on the
distribution of coronal magnetic energy density has been proposed to
quantitatively describe the CME's deflection in corona
\citep{Shen_etal_2010}, which points out that a CME will be
deflected towards a place with a lower energy density.

One may notice that there are some cross-equatorial deflections (the
red symbols in Fig.\ref{fg_def0}) and pole-ward deflections, which
have been listed in Table~\ref{tb_deflection}. Here the
cross-equatorial deflection is defined for the CMEs with $\Delta
PA-|Latitude|>20^\circ$, which we also called equator-ward
overshooting. The two kinds of deflections are not expected
according to the above analysis. A question is naturally raised
whether these unusual deflections are exceptional cases, or they can
also be described by the same model. According to the parameters
listed in Table~\ref{tb_deflection}, a quick impression can be
established for limb LI CMEs that all the pole-ward deflections
happened to narrow and slow CMEs, while all the equator-ward
overshootings were associated with wide and fast CMEs. The same
trend seemingly applies to AL CMEs. A further detailed study of all
the deflected CMEs including these unusual events will be pursued in
another paper.

\section{Concluding Remarks}\label{sec_conclusions}
By manually checking the LASCO and EIT movies of all the 1078
CMEs listed in CDAW CME catalog during 1997 -- 1998, the solar
surface sources of these CMEs are identified, and a web-based on-line
list of them with the information of their source locations is
established at \url{http://space.ustc.edu.cn/dreams/cme_sources}.
The source locations and apparent properties of CMEs have the following
features.

The distribution of CME source locations in latitude manifests a clear
bimodal appearance with two most probable peaks in $\pm(15^\circ-30^\circ)$,
which is consistent with the location of active region belt.
No CMEs came from polar regions (outside of $\pm75^\circ$). About 56\% of
detected CMEs occurred near the solar limb (refer to Sec.\ref{sec_dis}).
The average apparent speed of CMEs is about 435 km s$^{-1}$, and there
is no evident difference between the apparent speeds of on-disk and limb CMEs.
According to the analysis of limb CMEs, the average value of angular widths of CMEs is about
$59^\circ$, and about 65\% of them have a width from $30^\circ$ to $90^\circ$.
Generally, on-disk CMEs are twice wider than limb CMEs, which suggests a significant
projection effect (refer to Sec.\ref{sec_apparent}).

Further, through the analysis based on the source locations of all CMEs, we infer
many interesting results.
\begin{enumerate}
\item About 16\% of LASCO CMEs probably originated
from front-side solar disk but left no evident eruptive signatures in the FOV of
EIT, and a lower cut-off for the CME visibility in EIT, which corresponds
to the apparent speed range of 100 -- 200 km s$^{-1}$, probably exists. (Sec.\ref{sec_eit})

\item About 19\% of CMEs were not detected by LASCO, and the missing rate
has a trend to monotonomically increase as the CME source
location moves from limb to disk center. (Sec.\ref{sec_lasco})

\item About 32\% of front-side CMEs can not be recognized by SOHO, which becomes
a natural explanation of high rate of missing alert of geomagnetic storms and
is also in agreement with the previous results that about 18 -- 44\% of ICMEs do not
have the corresponding CMEs. (Sec.\ref{sec_invisible})

\item The brightness of a white-light CME at any heliocentric distance
is roughly positively-correlated with its speed,
which implies that (1) a bright transient recorded in white-light
coronagraphs is contributed by both a CME and the compressed
solar wind plasma surrounding the CME, and (2) the CME mass derived from
the brightness is probably overestimated. (Sec.\ref{sec_dis2})

\item Both projection effect and violent eruption are the major causes of
halo CMEs, but for limb halo CMEs, the latter should be the primary one. Overall,
there are about 25\% of halo CMEs stronger than the average level of CMEs. (Sec.\ref{sec_halo})

\item Most CMEs manifest deflection behaviors near the solar minimum.
About 62\% of CMEs underwent an equator-ward deflection with the average
deflection angle of $\sim22^\circ$, and about 5\% of CMEs had a significant
pole-ward deflection with the average angle of $\sim16^\circ$.
At high latitude regions (outside of $\pm45^\circ$), most CMEs
deflected towards equator. (Sec.\ref{sec_deflection})

\item The CME deflections can be classified into three types.
One is due to the asymmetrical expansion, one is the non-radial ejection, and
the other is the deflected propagation caused by the interaction of the CME
with other neighboring magnetic field structures. (Sec.\ref{sec_dclass})

\end{enumerate}

These findings help people understanding the CMEs viewed in white-light coronagraphs
more properly and precisely. We believe that some new and deeper
questions have emerged from these results. This paper presents our first work
established on the information of CME source locations, and gives us the overview of
white-light CMEs. In our follow-up works, we will continue to address issues,
e.g., the CME deflections, the role of active regions in producing CMEs, the
relationship of CMEs with flares, etc.

\acknowledgments{
We acknowledge the use of the data from
the CDAW CME catalog, which is generated and maintained at the
CDAW Data Center by NASA and The Catholic University of America
in cooperation with the Naval Research Laboratory. SOHO is a
project of international cooperation between ESA and NASA.
We thank J. Zhang at George Mason University for his advice
in identifying CMEs' source locations.
This research is supported by grants from 973 key project
2011CB811403, NSFC 40525014, 40874075, 40904046, FANEDD 200530, CAS KZCX2-YW-QN511,
100-Talent Program of CAS, and the fundamental research funds for the central universities.
}

\bibliographystyle{agufull}
\bibliography{../../ahareference}

\begin{thebibliography}{49}
\expandafter\ifx\csname natexlab\endcsname\relax\def\natexlab#1{#1}\fi

\bibitem[{{\it Andrews\/}(2002)}]{Andrews_2002}
Andrews, M.~D., The front-to-back asymmetry of coronal emission, {\it Sol.
  Phys.\/}, {\it 208\/}, 317--324, 2002.

\bibitem[{{\it Andrews\/}(2003)}]{Andrews_2003}
Andrews, M.~D., A search for {CMEs} associated with big flares, {\it Sol.
  Phys.\/}, {\it 218\/}, 261--279, 2003.

\bibitem[{{\it Burkepile et~al.\/}(2004){\it Burkepile, Hundhausen, Stanger,
  St.~Cyr, and Seiden\/}}]{Burkepile_etal_2004}
Burkepile, J.~T., A.~J. Hundhausen, A.~L. Stanger, O.~C. St.~Cyr, and J.~A.
  Seiden, Role of projection effects on solar coronal mass ejection properties:
  1. a study of {CMEs} associated with limb activity, {\it J. Geophys. Res.\/},
  {\it 109(A3)\/}, A03,103, 2004.

\bibitem[{{\it Cane and Richardson\/}(2003)}]{Cane_Richardson_2003}
Cane, H.~V., and I.~G. Richardson, Interplanetary coronal mass ejections in the
  near-earth solar wind during 1996--2002, {\it J. Geophys. Res.\/}, {\it
  108(A4)\/}, 1156, {doi:10.1029/2002JA009,817}, 2003.

\bibitem[{{\it Cane et~al.\/}(2000){\it Cane, Richardson, and {St.
  Cyr}\/}}]{Cane_etal_2000}
Cane, H.~V., I.~G. Richardson, and O.~C. {St. Cyr}, Coronal mass ejections,
  interplanetary ejecta and geomagnetic storms, {\it Geophys. Res. Lett.\/},
  {\it 27\/}, 3591--3594, 2000.

\bibitem[{{\it Cargill\/}(2004)}]{Cargill_etal_2004}
Cargill, P.~J., On the areodynamic drag force acting on interplanetary coronal
  mass ejections, {\it Sol. Phys.\/}, {\it 221\/}, 135--149, 2004.

\bibitem[{{\it Cargill et~al.\/}(1996){\it Cargill, Chen, Spicer, and
  Zalesak\/}}]{Cargill_etal_1996}
Cargill, P.~J., J.~Chen, D.~S. Spicer, and S.~T. Zalesak, Magnetohydrodynamic
  simulations of the motion of magnetic flux tubes through a magnetized plasma,
  {\it J. Geophys. Res.\/}, {\it 101\/}, 4855--4870, 1996.

\bibitem[{{\it Cremades and Bothmer\/}(2004)}]{Cremades_Bothmer_2004}
Cremades, H., and V.~Bothmer, On the three-dimensional configuration of coronal
  mass ejections, {\it Astron. \& Astrophys.\/}, {\it 422\/}, 307--322, 2004.

\bibitem[{{\it Cremades et~al.\/}(2006){\it Cremades, Bothmer, and
  Tripathi\/}}]{Cremades_etal_2006}
Cremades, H., V.~Bothmer, and D.~Tripathi, Properties of structured coronal
  mass ejections in solar cycle 23, {\it Adv. in Space Res.\/}, {\it 38\/},
  461--465, 2006.

\bibitem[{{\it Gopalswamy et~al.\/}(2000{\natexlab{a}}){\it Gopalswamy,
  Hanaoka, and Hudson\/}}]{Gopalswamy_etal_2000a}
Gopalswamy, N., Y.~Hanaoka, and H.~S. Hudson, Structure and dynamics of the
  corona surrounding an eruptive prominence, {\it Adv. in Space Res.\/}, {\it
  25\/}, 1851--1854, 2000{\natexlab{a}}.

\bibitem[{{\it Gopalswamy et~al.\/}(2000{\natexlab{b}}){\it Gopalswamy, Lara,
  Lepping, Kaiser, Berdichevsky, and {St. Cyr}\/}}]{Gopalswamy_etal_2000}
Gopalswamy, N., A.~Lara, R.~P. Lepping, M.~L. Kaiser, D.~Berdichevsky, and
  O.~C. {St. Cyr}, Interplanetary acceleration of coronal mass ejections, {\it
  Geophys. Res. Lett.\/}, {\it 27\/}, 145--148, 2000{\natexlab{b}}.

\bibitem[{{\it Gopalswamy et~al.\/}(2001){\it Gopalswamy, Yashiro, Kaiser,
  Howard, and Bougeret\/}}]{Gopalswamy_etal_2001c}
Gopalswamy, N., S.~Yashiro, M.~L. Kaiser, R.~A. Howard, and J.~L. Bougeret,
  Radio signatures of coronal mass ejection interaction: Coronal mass ejection
  cannibalism?, {\it Astrophys. J.\/}, {\it 548\/}, L91--L94, 2001.

\bibitem[{{\it Gopalswamy et~al.\/}(2004){\it Gopalswamy, Yashiro, Krucker,
  Stenborg, and Howard\/}}]{Gopalswamy_etal_2004}
Gopalswamy, N., S.~Yashiro, S.~Krucker, G.~Stenborg, and R.~A. Howard,
  Intensity variation of large solar energetic particle events associated with
  coronal mass ejections, {\it J. Geophys. Res.\/}, {\it 109\/}, A12,105, 2004.

\bibitem[{{\it Green et~al.\/}(2002){\it Green, Matthews, {van
  Driel-Gesztelyi}, Harra, and Culhane\/}}]{Green_etal_2002}
Green, L.~M., S.~A. Matthews, L.~{van Driel-Gesztelyi}, L.~K. Harra, and J.~L.
  Culhane, Multi-wavelength observations of an {X}-class flare without a
  coronal mass ejection, {\it Sol. Phys.\/}, {\it 205\/}, 325--339, 2002.

\bibitem[{{\it Harrison\/}(1995)}]{Harrison_1995}
Harrison, R.~A., The nature of solar flares associated with coronal mass
  ejection, {\it Astron. \& Astrophys.\/}, {\it 304\/}, 585--594, 1995.

\bibitem[{{\it Harrison\/}(2003)}]{Harrison_2003}
Harrison, R.~A., Soho observations relating to the association between flares
  and coronal mass ejections, {\it Adv. Space Res.\/}, {\it 32(12)\/},
  2425--2437, 2003.

\bibitem[{{\it Harrison and Lyons\/}(2000)}]{Harrison_Lyons_2000}
Harrison, R.~A., and M.~Lyons, A spectroscopic study of coronal dimming
  associated with a coronal mass ejection, {\it Astron. \& Astrophys.\/}, {\it
  358\/}, 1097--1108, 2000.

\bibitem[{{\it Howard et~al.\/}(1982){\it Howard, Michels, {Sheeley, Jr.}, and
  Koomen\/}}]{Howard_etal_1982}
Howard, R.~A., D.~J. Michels, N.~R. {Sheeley, Jr.}, and M.~J. Koomen, The
  observation of a coronal transient directed at earth, {\it Astrophys. J.\/},
  {\it 263\/}, L101--L104, 1982.

\bibitem[{{\it Hundhausen\/}(1993)}]{Hundhausen_1993}
Hundhausen, A.~J., Sizes and locations of coronal mass ejections -- {SMM}
  observations from 1980 and 1984--1989, {\it J. Geophys. Res.\/}, {\it
  98(A8)\/}, 13,177, 1993.

\bibitem[{{\it Lara et~al.\/}(2006){\it Lara, Gopalswamy, Xie, Mendoza-Torres,
  P\`{e}rez-Er\`{i}quez, and Michalek\/}}]{Lara_etal_2006}
Lara, A., N.~Gopalswamy, H.~Xie, E.~Mendoza-Torres, R.~P\`{e}rez-Er\`{i}quez,
  and G.~Michalek, Are halo coronal mass ejections special events?, {\it J.
  Geophys. Res.\/}, {\it 111\/}, A06,107, 2006.

\bibitem[{{\it Lindsay et~al.\/}(1999){\it Lindsay, Luhmann, Russell, and
  Gosling\/}}]{Lindsay_etal_1999}
Lindsay, G.~M., J.~G. Luhmann, C.~T. Russell, and J.~T. Gosling, Relationships
  between coronal mass ejection speeds from coronagraph images and
  interplanetary characteristics of associated interplanetary coronal mass
  ejections, {\it J. Geophys. Res.\/}, {\it 104\/}, 12,515, 1999.

\bibitem[{{\it MacQueen et~al.\/}(1986){\it MacQueen, Hundhausen, and
  Conover\/}}]{MacQueen_etal_1986}
MacQueen, R.~M., A.~J. Hundhausen, and C.~W. Conover, The propagation of
  coronal mass ejection transients, {\it J. Geophys. Res.\/}, {\it 91\/},
  31--38, 1986.

\bibitem[{{\it Plunkett et~al.\/}(2001){\it Plunkett, Thompson, {St. Cyr}, and
  Howard\/}}]{Plunkett_etal_2001}
Plunkett, S.~P., B.~J. Thompson, O.~C. {St. Cyr}, and R.~A. Howard, Solar
  source regions of coronal mass ejections and their geomagnetic effects, {\it
  J. Atmos. Solar-Terres. Phys.\/}, {\it 63\/}, 389--402, 2001.

\bibitem[{{\it Robbrecht et~al.\/}(2009){\it Robbrecht, Patsourakos, and
  Vourlidas\/}}]{Robbrecht_etal_2009}
Robbrecht, E., S.~Patsourakos, and A.~Vourlidas, No trace left behind: {STEREO}
  observation of a coronal mass ejection without low coronal signatures, {\it
  Astrophys. J.\/}, {\it 701\/}, 283--291, 2009.

\bibitem[{{\it Schwenn et~al.\/}(2005){\it Schwenn, {Dal Lago}, Huttunen, and
  Gonzalez\/}}]{Schwenn_etal_2005}
Schwenn, R., A.~{Dal Lago}, E.~Huttunen, and W.~D. Gonzalez, The association of
  coronal mass ejections with their effects near the {Earth}, {\it Annales
  Geophysicae\/}, {\it 23(3)\/}, 1033--1059, 2005.

\bibitem[{{\it Shen et~al.\/}(2010){\it Shen, Wang, Gui, Ye, and
  Wang\/}}]{Shen_etal_2010}
Shen, C., Y.~Wang, B.~Gui, P.~Ye, and S.~Wang, Kinematic evolution of a slow
  {CME} in near solar space viewed by {STEREO-B} in {October} 8, 2007, {\it
  Sol. Phys.\/}, {\it accepted\/}, 2010.

\bibitem[{{\it {St. Cyr} and Webb\/}(1991)}]{StCyr_Webb_1991}
{St. Cyr}, O.~C., and D.~F. Webb, Activity associated with coronal mass
  ejections at solar minimum - {SMM} observations from 1984-1986, {\it Sol.
  Phys.\/}, {\it 136\/}, 379--394, 1991.

\bibitem[{{\it {St. Cyr} et~al.\/}(2000){\it {St. Cyr}, Howard, {Sheeley, Jr.},
  Plunkett, Michels, Paswaters, Koomen, Simnett, Thompson, Gurman, Schwenn,
  Webb, Hildner, and Lamy\/}}]{StCyr_etal_2000}
{St. Cyr}, O.~C., R.~A. Howard, N.~R. {Sheeley, Jr.}, S.~P. Plunkett, D.~J.
  Michels, S.~E. Paswaters, M.~J. Koomen, G.~M. Simnett, B.~J. Thompson, J.~B.
  Gurman, R.~Schwenn, D.~F. Webb, E.~Hildner, and P.~L. Lamy, Properties of
  coronal mass ejections: {SOHO LASCO} observations from {January} 1996 to
  {June} 1998, {\it J. Geophys. Res.\/}, {\it 105\/}, 18,169, 2000.

\bibitem[{{\it Subramanian and Dere\/}(2001)}]{Subramanian_Dere_2001}
Subramanian, P., and K.~P. Dere, Source regions of coronal mass ejections, {\it
  Astrophys. J.\/}, {\it 561\/}, 372, 2001.

\bibitem[{{\it Vourlidas and Howard\/}(2006)}]{Vourlidas_Howard_2006}
Vourlidas, A., and R.~A. Howard, The proper treatment of coronal mass ejection
  brightness: A new methodology and implications for observations, {\it
  Astrophys. J.\/}, {\it 642\/}, 1216--1221, 2006.

\bibitem[{{\it Vourlidas et~al.\/}(2000){\it Vourlidas, Subramanian, Dere, and
  Howard\/}}]{Vourlidas_etal_2000}
Vourlidas, A., P.~Subramanian, K.~P. Dere, and R.~A. Howard, Large-angle
  spectrometric coronagraph measurements of the energetics of coronal mass
  ejections, {\it Astrophys. J.\/}, {\it 534\/}, 456--467, 2000.

\bibitem[{{\it Wang and Zhang\/}(2007)}]{Wang_Zhang_2007}
Wang, Y., and J.~Zhang, A comparative study between eruptive x-class flares
  associated with coronal mass ejections and confined x-class flares, {\it
  Astrophys. J.\/}, {\it 665\/}, 1428, 2007.

\bibitem[{{\it Wang et~al.\/}(2004){\it Wang, Shen, Ye, and
  Wang\/}}]{Wang_etal_2004b}
Wang, Y., C.~Shen, P.~Ye, and S.~Wang, Deflection of coronal mass ejection in
  the interplanetary medium, {\it Sol. Phys.\/}, {\it 222\/}, 329--343, 2004.

\bibitem[{{\it Wang et~al.\/}(2006){\it Wang, Xue, Shen, Ye, Wang, and
  Zhang\/}}]{Wang_etal_2006a}
Wang, Y., X.~Xue, C.~Shen, P.~Ye, S.~Wang, and J.~Zhang, Impact of the major
  coronal mass ejections on geo-space during {September} 7 -- 13, 2005, {\it
  Astrophys. J.\/}, {\it 646\/}, 625--633, 2006.

\bibitem[{{\it Wang et~al.\/}(2002){\it Wang, Ye, Wang, Zhou, and
  Wang\/}}]{Wang_etal_2002a}
Wang, Y.~M., P.~Z. Ye, S.~Wang, G.~P. Zhou, and J.~X. Wang, A statistical study
  on the geoeffectiveness of earth-directed coronal mass ejections from {March}
  1997 to {December} 2000, {\it J. Geophys. Res.\/}, {\it 107(A11)\/}, 1340,
  doi:10.1029/2002JA009,244, 2002.

\bibitem[{{\it Webb et~al.\/}(1997){\it Webb, Kahler, McIntosh, and
  Klimchuck\/}}]{Webb_etal_1997}
Webb, D.~F., S.~W. Kahler, P.~S. McIntosh, and J.~A. Klimchuck, Large-scale
  structures and multiple neutral lines associated with coronal mass ejections,
  {\it J. Geophys. Res.\/}, {\it 102(A11)\/}, 24,161--24,174, 1997.

\bibitem[{{\it Webb et~al.\/}(1998){\it Webb, Cliver, Gopalswamy, Hudson, and
  {St. Cyr}\/}}]{Webb_etal_1998}
Webb, D.~F., E.~W. Cliver, N.~Gopalswamy, H.~S. Hudson, and O.~C. {St. Cyr},
  The solar origin of the january 1997 coronal mass ejection, magnetic cloud
  and geomagnetic storm, {\it Geophys. Res. Lett.\/}, {\it 25\/}, 2469--2472,
  1998.

\bibitem[{{\it Webb et~al.\/}(2000){\it Webb, Cliver, Crooker, {St. Cyr}, and
  Thompson\/}}]{Webb_etal_2000}
Webb, D.~F., E.~W. Cliver, N.~U. Crooker, O.~C. {St. Cyr}, and B.~J. Thompson,
  Relationship of halo coronal mass ejections, magnetic clouds, and magnetic
  storms, {\it J. Geophys. Res.\/}, {\it 105(A4)\/}, 7491--7508, 2000.

\bibitem[{{\it Xiong et~al.\/}(2006){\it Xiong, Zheng, Wang, and
  Wang\/}}]{Xiong_etal_2006}
Xiong, M., H.~Zheng, Y.~Wang, and S.~Wang, Magnetohydrodynamic simulation of
  the interaction between interplanetary strong shock and magnetic cloud and
  its consequent geoeffectiveness: 2. oblique collision, {\it J. Geophys.
  Res.\/}, {\it 111\/}, A11,102, 2006.

\bibitem[{{\it Xiong et~al.\/}(2009){\it Xiong, Zheng, and
  Wang\/}}]{Xiong_etal_2009}
Xiong, M., H.~Zheng, and S.~Wang, Magnetohydrodynamic simulation of the
  interaction between two interplanetary magnetic clouds and its consequent
  geoeffectiveness: 2. oblique collision, {\it J. Geophys. Res.\/}, {\it
  114\/}, A11,101, 2009.

\bibitem[{{\it Yashiro et~al.\/}(2004){\it Yashiro, Gopalswamy, Michalek, {St.
  Cyr}, Plunkett, Rich, and Howard\/}}]{Yashiro_etal_2004}
Yashiro, S., N.~Gopalswamy, G.~Michalek, O.~C. {St. Cyr}, S.~P. Plunkett, N.~B.
  Rich, and R.~A. Howard, A catalog of white light coronal mass ejections
  observed by the soho spacecraft, {\it J. Geophys. Res.\/}, {\it 109(A7)\/},
  A07,105, 2004.

\bibitem[{{\it Yashiro et~al.\/}(2005){\it Yashiro, Gopalswamy, Akiyama,
  Michalek, and Howard\/}}]{Yashiro_etal_2005}
Yashiro, S., N.~Gopalswamy, S.~Akiyama, G.~Michalek, and R.~A. Howard,
  Visibility of coronal mass ejections as a function of flare location and
  intensity, {\it J. Geophys. Res.\/}, {\it 110(A12)\/}, A12S05, 2005.

\bibitem[{{\it Yermolaev\/}(2008)}]{Yermolaev_2008}
Yermolaev, Y.~I., Comment on ``geoeffectiveness of halo coronal mass
  ejections'' by n. gopalswamy, s. yashiro, and s. akiyama (j. geophys. res.
  2007, 112, doi:10.1029/2006ja012149), {\it Cosmic Research\/}, {\it 46\/},
  540--541, 2008.

\bibitem[{{\it Yermolaev and Yermolaev\/}(2006)}]{Yermolaev_Yermolaev_2006}
Yermolaev, Y.~I., and M.~Yermolaev, Statistic study on the geomagnetic storm
  effectiveness of solar and interplanetary events, {\it Adv. in Space Res.\/},
  {\it 37\/}, 1175--1181, 2006.

\bibitem[{{\it Zhang et~al.\/}(2001){\it Zhang, Dere, Howard, Kundu, and
  White\/}}]{Zhang_etal_2001a}
Zhang, J., K.~P. Dere, R.~A. Howard, M.~R. Kundu, and S.~M. White, On the
  temporal relationship between coronal mass ejections and flares, {\it
  Astrophys. J.\/}, {\it 559\/}, 452--462, 2001.

\bibitem[{{\it Zhang et~al.\/}(2003){\it Zhang, Dere, Howard, and
  Bothmer\/}}]{Zhang_etal_2003}
Zhang, J., K.~P. Dere, R.~A. Howard, and V.~Bothmer, Identification of solar
  sources of major geomagnetic storms between 1996 and 2000, {\it Astrophys.
  J.\/}, {\it 582\/}, 520--533, 2003.

\bibitem[{{\it Zhang et~al.\/}(2007){\it Zhang, Richardson, Webb, Gopalswamy,
  Huttunen, Kasper, Nitta, Poomvises, Thompson, Wu, Yashiro, and
  Zhukov\/}}]{Zhang_etal_2007}
Zhang, J., I.~G. Richardson, D.~F. Webb, N.~Gopalswamy, E.~Huttunen, J.~C.
  Kasper, N.~V. Nitta, W.~Poomvises, B.~J. Thompson, C.-C. Wu, S.~Yashiro, and
  A.~N. Zhukov, Solar and interplanetary sources of major geomagnetic storms
  ({Dst} $\leq$ -100 {nT}) during 1996-2005, {\it J. Geophys. Res.\/}, {\it
  112\/}, A10,102, 2007.

\bibitem[{{\it Zhao and Webb\/}(2003)}]{Zhao_Webb_2003}
Zhao, X.~P., and D.~F. Webb, Source regions and storm effectiveness of
  frontside full halo coronal mass ejections, {\it J. Geophys. Res.\/}, {\it
  108(A6)\/}, 1234, 2003.

\bibitem[{{\it Zhou et~al.\/}(2003){\it Zhou, Wang, and
  Cao\/}}]{Zhou_etal_2003}
Zhou, G., J.~Wang, and Z.~Cao, Correlation between halo coronal mass ejections
  and solar surface activity, {\it Astron. \& Astrophys.\/}, {\it 397\/}, 1057,
  2003.

\end{thebibliography}

\end{document}